\documentclass[preprint]{aastex}
\usepackage{apjfonts}[11pt]

\begin{document}

\title{ MERGING GALAXY CLUSTER ABELL 2255 IN MID-INFRARED }

\author{Hyunjin Shim\altaffilmark{1,2,3}, Myungshin Im\altaffilmark{2,3},
 Hyung Mok Lee\altaffilmark{2},  Myung Gyoon Lee\altaffilmark{2},
 Seong Jin Kim\altaffilmark{2}, 
 Ho Seong Hwang\altaffilmark{4}, Narae Hwang\altaffilmark{5},
 Jongwan Ko\altaffilmark{2,3}, Jong Chul Lee\altaffilmark{2},
 Sungsoon Lim\altaffilmark{2}, Hideo Matsuhara\altaffilmark{6},
 Hyunjong Seo\altaffilmark{2}, Takehiko Wada\altaffilmark{6},
 and Tomotsugu Goto\altaffilmark{7}
 }

\altaffiltext{1}{Spitzer Science Center, California Institute of Technology,
 MS 220-6, Pasadena, CA 91125 ; hjshim@ipac.caltech.edu }
\altaffiltext{2}{Astronomy Program, Department of Physics \& Astronomy, FPRD,
 Seoul National University, Seoul, Korea }
\altaffiltext{3}{Center for the Exploration of the Origin of the Universe
 (CEOU), Seoul National University, Seoul, Korea }
\altaffiltext{4}{CEA Saclay/Service d'Astrophysique, F-91191 Gif-sur-Yvette, France}
\altaffiltext{5}{National Astronomical Observatory of Japan,
 Mitaka, Tokyo 181-8588, Japan}
\altaffiltext{6}{Institute of Space and Astronautical Science,
 Japan Aerospace Exploration Agency, Kanagawa 229-8510, Japan}
\altaffiltext{7}{Institute for Astronomy, 
 University of Hawaii, 2680 Woodlawn Drive, Honolulu, HT 96822}

\begin{abstract}
  We present the mid-infrared (MIR) observation of a nearby
 galaxy cluster, Abell 2255 by the AKARI space telescope. Using
 the AKARI's continuous wavelength coverage between 3-24\,$\mu$m 
 and the wide field of view, we investigate the properties of
 cluster member galaxies to see how the infall of the galaxies, 
 the cluster substructures, and the cluster-cluster merger influence
 their evolution.
  We show that the excess of MIR ($\sim11~\mu$m) flux is a good
 indicator to discriminate galaxies at different evolutionary
 stages, and divide galaxies into three classes accordingly :
 strong MIR-excess ($N3-S11>0.2$) galaxies that include both
 unobscured and obscured star-forming galaxies, weak MIR-excess 
 ($-2.0<N3-S11<-1.2$) galaxies that are quiescent, old ($>5$\,Gyr)
 galaxies where the MIR emission arises mainly from the circumstellar 
 dust around AGB stars,
  and intermediate MIR-excess ($-1.2<N3-S11<0.2$)
 galaxies in between the two classes
 that are less than a few Gyrs old past the
 prime star formation activity.
  With the MIR-excess diagnostics, we investigate how local 
 and cluster-scale environments affect the individual galaxies.
  We derive the total star formation rate and the specific 
 star formation rate of A2255 using the strong MIR-excess galaxies.
 The dust-free, total star formation rate of A2255 is 
 $\sim130~M_{\odot}$ yr$^{-1}$, which is consistent with the
 star formation rates of other clusters of galaxies at similar
 redshifts and with similar masses. We find no strong evidence 
 that supports enhanced star formation neither inside the cluster 
 nor in the substructure region, suggesting that the infall or
 the cluster merging activities tend to suppress the star formation.  
  The intermediate MIR-excess galaxies, representing galaxies in 
 transition from star-forming galaxies to quiescent galaxies, 
 are located preferentially at the medium density region 
 or cluster substructures with higher surface density of galaxies. 
  Our findings suggest that galaxies are being transformed 
 from star-forming galaxies into red, quiescent galaxies 
 from the infall region through near the core which can be 
 well-explained by the ram-pressure stripping as previous simulation
 results suggest. 
  We conclude that the cluster merging and the group/galaxy infall 
 suppress the star formation and transform galaxies from star-forming
 galaxies into quiescent galaxies, most likely due to ram-pressure stripping. 
\end{abstract}


\section{Introduction}

  The formation and the evolution of galaxies are thought to be 
 strongly dependent on their environments.
  As an extreme example of high-density environment,
 galaxy clusters may affect star formation activities of their
 member galaxies through various processes
 (Boselli \& Gavazzi 2006; Park \& Hwang 2009),
 including ram-pressure stripping
 (Gunn \& Gott 1972; Abadi, Moore, \& Bower 1999),
 cluster tidal forces (Fujita 1998),
 violent galaxy encounters (Lavery \& Henry 1994),
 or rapid galaxy encounters such as galaxy harassment (Moore et al. 1996). 
  These mechanisms are attributed as the reasons why galaxy clusters 
 show less star formation than field both at low and high 
 redshifts (e.g., Dressler \& Gunn 1983; Poggianti et al. 1999, 2009).
  
  To date, a number of galaxy clusters are known to be in a merging process 
 by their asymmetric X-ray emission and substructures in member 
 distribution (e.g., Markevitch et al. 1998; Donnelly et al. 2001).  
 The effect of cluster-scale merging on the member galaxies is
 still being debated (e.g., Hwang \& Lee 2009) : the cluster-cluster merging can
 trigger the star formation in member galaxies by driving the external potential
 (e.g., Bekki 1999), or quench the star formation by depriving gas
 via ram pressure of the intracluster medium (e.g.,
 Fujita et al. 1999). Both hypotheses are supported by observations,
 thus so far it is unclear which case is more dominant. 

   Abell 2255 (hereafter A2255) is a rich galaxy cluster at low
 redshift ($z=0.0806$; Struble \& Rood 1999) that consists of a few hundred
 member galaxies. Due to its richness (richness class 2; Abell 1958)
 and the advantage of extensive membership studies 
 (cluster membership is determined partly using the spectroscopic
 redshifts from the Sloan Digital Sky Survey Data Release 2, Abazajian et al. 2004;
 and partly using the photometric redshifts by Yuan, Zhou, \& Jiang 2003), 
 the cluster is ideal for investigation of stellar populations and evolution
 of its member galaxies.
   Previous studies based on various wavelengths -- X-ray, optical, and radio --
 have suggested that A2255 is not a relaxed cluster, but in a process of 
 cluster-cluster merging. The X-ray contour map is elongated in the east-west
 direction which implies that there was a merger in this direction 
 (Davis \& White 1998; Feretti et al. 1997).
 Since there are no distinct two peaks in the X-ray temperature distribution,
 it seems that A2255 is already in the late stage of cluster-cluster merger. 
 Optical member identification -- using spectroscopic and photometric redshifts 
 -- shows that A2255 has several significant substructures with different 
 velocity components \citep{Yuan03}. These substructures are interpreted
 as groups of galaxies infalling into the main cluster.
 Radio observations revealed an excess of radio galaxies in A2255 \citep{MO03},
 some of them being AGN-dominated.
 \cite{DMM03} found X-ray point sources in A2255 that are thought to be AGNs,
 and they also found that the total number of AGNs in A2255 is significantly
 larger than that of a typical galaxy cluster. 
 Finally, a possible alignment of star-forming galaxies that prefer a specific
 direction (south-north; Miller \& Owen 2003) can also be interpreted as the
 result of cluster-cluster merger.
 All these observations support the idea that A2255 is a good example of 
 an ``unrelaxed'' galaxy cluster.

  Yuan et al.(2005) discussed the effects of cluster-cluster merging  
 on the evolution of member galaxies using the SDSS spectroscopy.
  From the morphological analysis, they suggested that the cluster-cluster
 merger has different effects on the star formation activity of galaxies
 with different morphologies : star formation activity is suppressed in 
 E/S0 galaxies and enhanced in spiral and irregular galaxies.

  However, this previous study on the star formation of member galaxies is limited
 to the optically bright galaxies with spectroscopic information. 
 It is not clear whether these optically bright galaxies account for the
 majority of the star formation taking place in A2255.
  Firstly, there could be member galaxies that contribute significantly 
 to the total ongoing star formation despite their faint magnitudes in optical. 
  Secondly, there could be a heavily 
 obscured star formation that is not accounted for
 with the extinction correction estimated from the optical emission line ratios, 
 as witnessed by recent infrared (IR) studies of the obscured star formation
 in the core of several clusters of galaxies (e.g., Bai et al. 2007; 
 Marcillac et al. 2007).
  The mid-infrared (MIR) emission is an efficient probes of these 
 ``hidden'' star formation, since it is proportional to the total 
 IR luminosity (e.g., Chary \& Elbaz 2001). The MIR emission between 
 3-10\,$\mu$m also plays an important role in detecting the existence of
 dust inside galaxies (Bressan et al. 2007), which provides hints to the 
 evolution of galaxies.

  In this paper, we present 3-24\,$\mu$m observation
 of $\sim1200$ arcmin$^2$ over A2255 field using the IR telescope AKARI.
 By combining AKARI MIR data
 with the optical/X-ray/radio data and \textit{Spitzer}
 24/70\,$\mu$m data, we study the properties of cluster member galaxies
 to investigate the effect of cluster-cluster merger on their evolution.
  In particular, we show that a significant fraction of cluster member galaxies
 have ``excess'' in MIR with respect to the photospheric emission, i.e. 
 emission from stellar photosphere. The MIR-excess proves the presence
 of dust emission in these galaxies, which is an important sign of galaxy
 evolution. Thus, we focus on these ``MIR-excess'' galaxies and investigate
 their properties in relation to cluster dynamics, including cluster-cluster
 merger and infall. 
 
  We present the MIR and other ancillary data used in this study in Section 2. 
 In Section 3, we describe how MIR properties (colors) reflect the properties
 of A2255 member galaxies, and define the MIR diagnostics to study 
 galaxies of different levels of MIR-excess. 
  Using the MIR diagnostics, we derive the star formation rates of
 individual galaxies and the entire cluster in Section 4. In Section 5,
 we discuss the relation between the environmental parameters 
 and the MIR-excess galaxies. More detailed discussion suggesting the
 scenario of cluster-scale merging in A2255 is followed in Sections 6 and 7.
  Throughout this paper, we use a cosmology with 
 $\Omega_{M}=0.3$, $\Omega_{\Lambda}=0.7$, and $H_0=70$\,km\,s$^{-1}$\,Mpc$^{-1}$
 (e.g., Im, Griffiths, \& Ratnatunga 1997).
 All the magnitudes are given in AB system.

\section{Data}

 \subsection{AKARI Observation and Data Reduction}

  CLEVL (CLusters of galaxies EVoLution studies) is one of the
 Mission Programs of the AKARI IR telescope 
 (Murakami et al. 2007)
 designed to understand the formation and the evolution of galaxies
 in cluster environments.
  The program is divided into three major components according to
 the redshifts of the targets -- low-redshift, intermediate-redshift,
 and high-redshift clusters of galaxies. A2255 is one of the target
 galaxy clusters for the low-redshift CLEVL program 
 (Im et al. 2008; Lee et al. 2010).

\begin{figure}
 \epsscale{1.0}\plotone{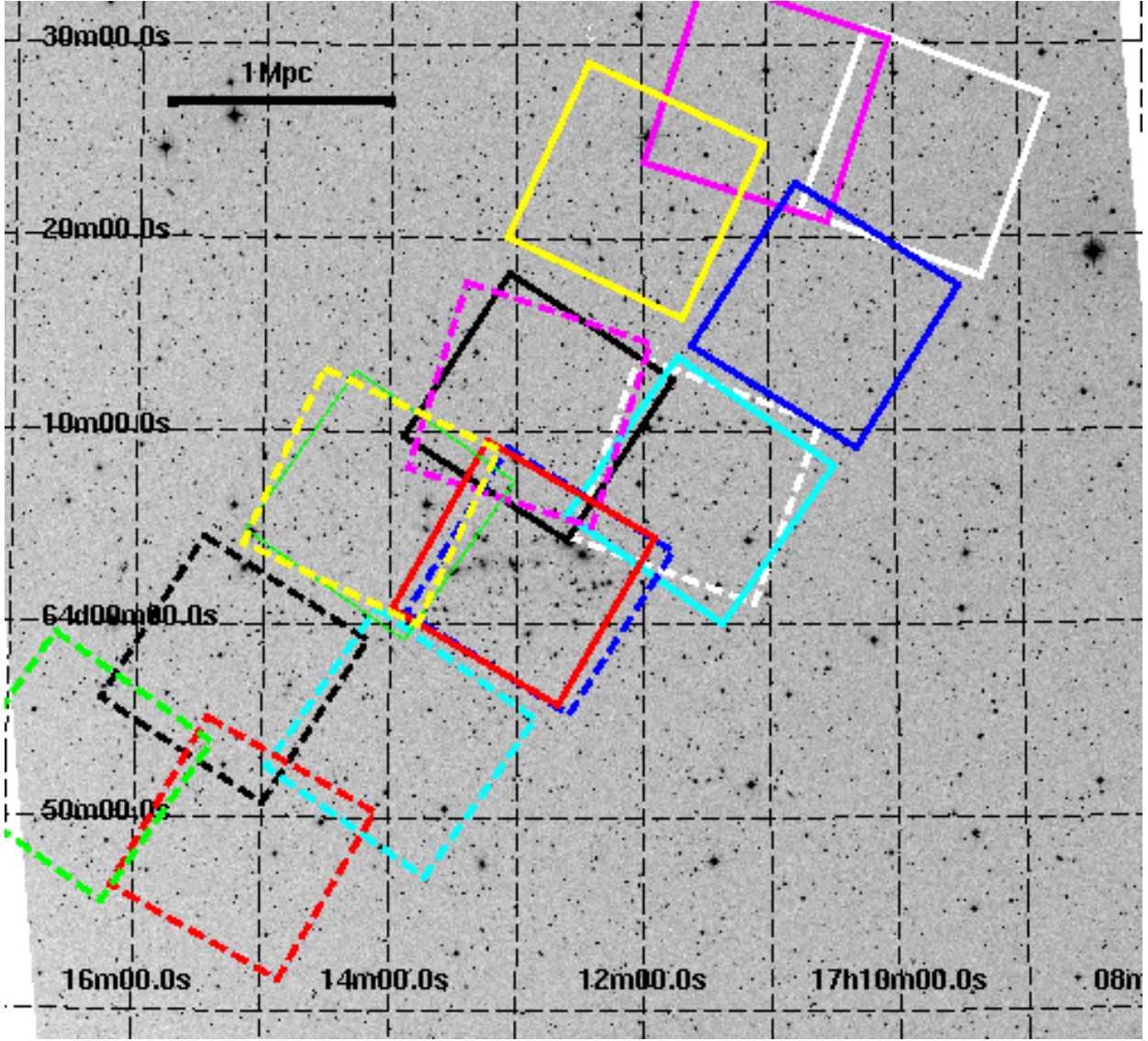}
 \caption{\label{fig:pointing} The field of views observed by the AKARI
 overlaid on the DSS optical image of A2255. In total, twelve 
 $10\times10$ arcmin$^2$ fields are observed, among which a pair of fields
 marked with the same color are observed simultaneously with the MIR-L and 
 NIR/MIR-S cameras. The \textit{dashed} boxes indicate
 the MIR-L camera fields (L15 and L24), while the \textit{solid} boxes
 indicate the regions observed by NIR/MIR-S camera (N3, N4, S7, and S11).
 }
 \epsscale{1.0}
\end{figure}

  A2255 is observed by AKARI InfraRed Camera (IRC; Onaka et al. 2007),
 over eight fields (Figure \ref{fig:pointing}) 
 with each field covering an area of $10\times10$ arcmin$^2$.
 The IRC consists of three cameras --
 NIR band camera [ $N2, N3, N4$ ], MIR-S band camera [ $S7, S9W, S11$ ],
 and MIR-L band camera [ $L15, L18W, L24$ ], where the number next to 
 each alphabet denotes the central wavelength of the filter. Among these,
 NIR and MIR-S band cameras are placed to observe the same field of view
 (FOV) on the sky, while MIR-L band camera points to
 a different field separated by $\sim10$ arcmin from the NIR and MIR-S
 field of view. 
  The wide field of view ($10\times10$ arcmin$^2$) and the continuous
 wavelength coverage at 3-24\,$\mu$m are two main advantages of AKARI
 with respect to \textit{Spitzer}, in terms of study of galaxy clusters.
  For the CLEVL low-redshift cluster programs, we used \texttt{IRC02}
 Astronomical Observation Template (AOT; see AKARI Observer's Manual ver
 1.2\footnote{http://www.ir.isas.jaxa.jp/ASTRO-F/Observation/ObsMan/akobsman12.pdf})  
 mode. This AOT takes moderate-length exposure by using two filters for each camera.
 The filter composition we selected is [ $N3, N4, S7, S11, L15$ \& $L24$ ]. 
  With this observational design, we obtain all-filter coverage for 
 the central four fields ($\sim400$\,arcmin$^2$).
  Four fields in the south are covered only in $L15$ and $L24$
 (dashed line in Figure \ref{fig:pointing}), while four fields in the north
 are covered in $N3/N4/S7/S11$ (solid line in Figure \ref{fig:pointing}).
  Thus, the final data coverage with at least one AKARI filter is 1200\,arcmin$^2$.
  The on-source exposure time for each filter is roughly $\sim140$\,seconds.
  The summary of the observation and the characteristics of the reduced
 images are specified in Table \ref{tab:obs_summary}. 

  The images are reduced using the IRC pipeline version 070104 
 (Y. Ita, T. Wada et al. 2007\footnote
 {http://www.ir.isas.jaxa.jp/ASTRO-F/Observation/DataReduction/IRC/software/irc20070104.tgz};
 provided as a form of IRAF external package). 
 The IRC pipeline consists of sky subtraction,
 astrometric calibration and final coaddition of individual frames.
  For the MIR-L images whose astrometric calibration within the pipeline is
 relatively poor, we derive an astrometric solution using the IRAF task
 \texttt{ccmap} using 2MASS sources or $S7$ images as a reference. 
  The full-width-at-half-maximum (FWHM) of the reduced images is between
 4.0-$6.0\arcsec$, depending on the filter. The astrometric accuracy
 is 1-$2\arcsec$ rms for $N3/N4/S7/S11$ images, and 3-$4\arcsec$ rms 
 for $L15/L24$ images.

  To measure the MIR fluxes, we use \textit{SExtractor} 
 (Bertin \& Arnouts 1996). In order to include all objects detected in 
 at least one band, we make a reference image for source detection by 
 combining images of different bands : 
  for the photometry of $N3$ through $S11$ images, 
 we use a $N3/N4/S7/S11$ combined image as a reference
 image. Since the sensitivities in $L15$ and $L24$ images are relatively
 poor compared to the case of $N3$-$S11$, we only combine L15 and L24 
 images to make a reference image for the photometry in $L15/L24$ bands.
 We perform dual-mode photometry using the reference images, with 
 \textit{SExtractor} configuration of \texttt{DETECT\_THRESH}$=3.0$,
 \texttt{DETECT\_MINAREA}$=2.0$, and \texttt{BACKGROUND} mesh size of
 32 pixels. 
  As a measure of the total flux of each galaxy, we use
 \texttt{FLUX\_AUTO} from \textit{SExtractor} output,
 i.e. the flux within a Kron elliptical aperture. The fluxes with a unit
 of ADU are converted to $f_{\nu}$ in $\mu$Jy, using the IRC flux calibration
 table in the AKARI IRC data users' manual ver 1.4
 (08/06/03; Lorente et al. 2007)\footnote
 {http://www.ir.isas.jaxa.jp/ASTRO-F/Observation/IDUM/IRC\_IDUM\_1.4.pdf}.
 The ``dual-mode'' photometry we used in this paper is consistent with a
 ``single-mode'' photometry, which is a photometry performed on
 each image independently. The difference between the 
 dual-mode photometry and the single-mode photometry is at most 
 $<5$\% for each object.

 \subsection{Cluster Member Identification}

  After the construction of the AKARI band-merged catalog, we matched
 the AKARI photometry catalog with catalogs of previously known
 cluster member galaxies. In order to identify the membership for 
 A2255, we use the spectroscopic redshifts from the SDSS Data Release 2
 (SDSS DR2; Abazajian et al. 2004) and BATC photometric redshifts by 
 \cite{Yuan03}. 
  Yuan, Zhou, \& Jiang (2003) constructed a cluster member catalog
 that consists of 214 spectroscopic members and 313 photometric members 
 out to $\sim3$\,Mpc from the cluster center.
  Since all the AKARI sources fall within the
 coverage of this catalog, we use Yuan, Zhou, \& Jiang (2003)'s 
 cluster member catalog 
 (hereafter the Y03 catalog) as a basis of our analysis in the following 
 sections. 

  The SDSS spectroscopic targets are selected down to $r\lesssim17.77$
 mag after correcting for Galactic extinction (Strauss et al. 2002),
 therefore all members with spectroscopic redshifts are brighter than 
 17.77 mag in $r$-band. The completeness of the cluster member selection
 with spectroscopic redshifts at this magnitude range is on average
 $\sim85$\%, although it depends on the magnitudes and the clustercentric
 distances of the galaxies (see Fig. 1 in Park \& Hwang 2009). 
  The addition of cluster members with photometric redshifts corrects
 this incompleteness. Note that the number of member galaxies with 
 photometric redshifts brighter than $r=17.77$ mag is 40. 
  The magnitude limit for the BATC photometry is
 $\sim20$ mag in each BATC filter, and the BATC photometric redshift
 identification is nearly complete for galaxies brighter than $V<19$ mag
 (Yuan, Zhou, \& Jiang 2003). This magnitude corresponds to 
 $r<18.6$-18.9 mag for different types of galaxies (c.f. galaxy colors
 in various photometric band systems; Fukugita, Shimasaku, \& Ichikawa 1995).
  Therefore, the cluster member selection is complete down to 
 $r<18.6$-18.9 mag in terms of both the redshift identification and 
 the member identification. 
  Above this limit, there are more than $\sim100$ member galaxies with 
 $18.9<r<20$ mag. We expect a significant incompleteness
 in cluster member selection for this magnitude range. Although it is
 difficult to estimate the exact incompleteness at this faint end, 
 this sample incompleteness would not cause much problem considering
 the relatively high MIR detection limit compared to the optical limit.
  We describe the effect of the incompleteness on our analysis 
 in more detail in Section 4 (also, please refer to
 Figure \ref{fig:irlimit}).
  The redshift cut for A2255 member galaxies is $0.068<z<0.090$,
 and the median redshift of A2255 member galaxies is 
 $\langle z \rangle=0.081$.

  The matching between the Y03 catalog and the AKARI IR source catalog is
 done using a matching radius of $5\arcsec$ (roughly $\sim1.5\times$FWHM
 radius of the AKARI N3 image, close to FWHM in other bands). We checked the
 optical and AKARI images of individual member galaxy to confirm that
 the matched object is not blended or mis-matched with neighboring sources.
 Most ($\sim95$\%) of
 the member galaxies are not blended in MIR. For 5\% of galaxies that
 are suspected to be blended in MIR, we assign `$-1.00$' value
 in the final photometry table (Table \ref{tab:phot}). Galaxies outside
 the survey coverage are also assigned `$-1.00$', and these are not
 included in following analysis. If the galaxy is within the survey 
 coverage and has the flux below the detection limit, we assign `$99.00$'.
 We identify 122 spectroscopically
 confirmed member galaxies and 170 member galaxies selected by 
 photometric redshifts (292 in total) in the Y03 catalog which have 
 AKARI MIR counterparts, over $\sim1200$\,arcmin$^2$ of the AKARI observation.

 \subsection{\textit{Spitzer} MIPS Photometry}

  A2255 is observed by \textit{Spitzer} MIPS 24\,$\mu$m and 70\,$\mu$m at 
 $5\sigma$ flux limits of $250~\mu$Jy and 5\,mJy respectively (PID 40562, 
 PI: G. Rieke). Since the coverage of MIPS 24\,$\mu$m is $\sim1500$\,arcmin$^2$,
 and it includes the region not covered by AKARI (see Figure \ref{fig:substructure}b
 for comparison of the AKARI IRC and the \textit{Spitzer} MIPS coverages),
 we derive the MIPS 24/70\,$\mu$m photometry for cluster member galaxies
 within the \textit{Spitzer} MIPS coverage.
  The inclusion of 24/70\,$\mu$m-detected galaxies 
 increases the number of member galaxies detected in MIR.

  We run \textit{SExtractor} over \texttt{pbcd} (post-Basic Calibrated Data) 
 MIPS images to measure 24/70\,$\mu$m fluxes. We use \texttt{FLUX\_AUTO} 
 of each object as total flux like we did in case of AKARI IRC bands.
 The \texttt{FLUX\_AUTO} is measured at the previously defined
 optical coordinates of galaxies (i.e. the Y03 catalog)
 using \texttt{ASSOC} parameters. All AKARI-Y03 matched members have either
 24/70\,$\mu$m flux or upper limits, and there are 134 additional member 
 galaxies in the Y03 catalog that lies within \textit{Spitzer} MIPS coverage
 and outside the AKARI coverage. 

  In Table \ref{tab:phot}, we present MIR photometry of A2255 member galaxies
 from the AKARI and the \textit{Spitzer} observations. The table contains
 photometry for 426 member galaxies from the Y03 catalog : 
 292 galaxies within AKARI IRC fields, 134 galaxies within \textit{Spitzer}
 MIPS coverage but outside the AKARI IRC coverage. The columns in the table
 are -- galaxy id (with coordinate information),
 spectroscopic or photometric redshift, AKARI IRC flux (3, 4, 7,
 11, 15, and 24\,$\mu$m), and MIPS 24 and 70\,$\mu$m flux. As mentioned in 
 Section 2.2, the value of $-1.00$ in the magnitude column indicates that the
 object is not covered by the corresponding filter, and $99.00$ indicates
 that the flux of the object is below the detection limit.

 \subsection{ Multi-Wavelength Ancillary Data }

  In this section, we describe other wavelength data used in this 
 study : optical imaging and spectroscopy, UV imaging, radio, and X-ray 
 imaging. 

  In the optical wavelengths, we use the $ugriz$ SDSS photometry
 as well as the 13-band photometry (from UV to $i$-band) in Yuan, Zhou, \&
 Jiang (2003). These optical photometric points are used to construct the 
 spectral energy distributions of cluster member galaxies (see Section 3.3
 for details). We also use the spectroscopic redshift from the SDSS 
 database for the membership identification. The line equivalent 
 widths, and the stellar metallicities are taken from the SDSS-MPA 
 catalog\footnote{http://www.mpa-garching.mpg.de/SDSS/DR4/} and 
 from Gallazzi et al. (2005).

  We also used the UV flux information which provides
 another measure of SFRs 
 (see Section 4.1 for details). We queried the GALEX source catalog
 from the all sky survey and 
 the nearby galaxies survey\footnote{http://galex.stsci.edu/GR4/}
 (tile \texttt{UVE\_A2255}) over $\sim1$ deg$^2$ of A2255. 
  We found 313 matches within $10\arcsec$ from the optical coordinates
 of A2255 member galaxies within the MIR coverage 
 (Note the total number of member galaxies with the MIR data is 426; 
 Section 2.3).
  For 76 galaxies that are used in the comparison between UV- and IR-
 derived SFRs (Section 4.1), we visually inspected the UV images 
 to confirm whether the cross-identification is correct and the 
 UV flux is not contaminated by nearby sources.
  The depths in FUV ($1500 \mbox{\AA}$)
 and NUV ($2300 \mbox{\AA}$) wavelengths are roughly $\sim20~\mu$Jy at 
 $5~\sigma$ flux limit. 

  The radio flux is also used in the comparison between the 
 different SFR indicators (Section 4.1).
  We use 1.4 GHz radio source catalog
 of Miller \& Owen (2003) and cross-identify the sources with IR-detected
 cluster members using a matching radius of $5\arcsec$. 
  The number of matched members within MIR field of view is 33.
 Among these, 20 galaxies are classified as star-forming galaxies,
 three as Seyfert galaxies, 
 and the rest ten galaxies as AGN candidates with old stellar
 population (Miller \& Owen 2003; the classification is based on the
 optical spectra of galaxies). 

  Finally, we use the X-ray point source catalog obtained with Chandra  
 (Davis, Miller, \& Mushotzky 2003) and cross-identify 8 sources using
 a matching radius of $5\arcsec$. These sources overlap with radio 
 sources (Miller \& Owen 2003) described above. We present a separate
 section about AGN in Section 3.5.

\section{MIR Photometric Properties of A2255 Galaxies and MIR-Excess Galaxies}

  In this section, we explore the MIR properties of galaxies in A2255, 
 and provide a classification method based on the excess in the MIR 
 emission in order to facilitate the study of environmental dependence 
 of the galaxy evolution in the cluster.

 \subsection{Color-Magnitude Relation}

\begin{figure}
\epsscale{0.75}
\plotone{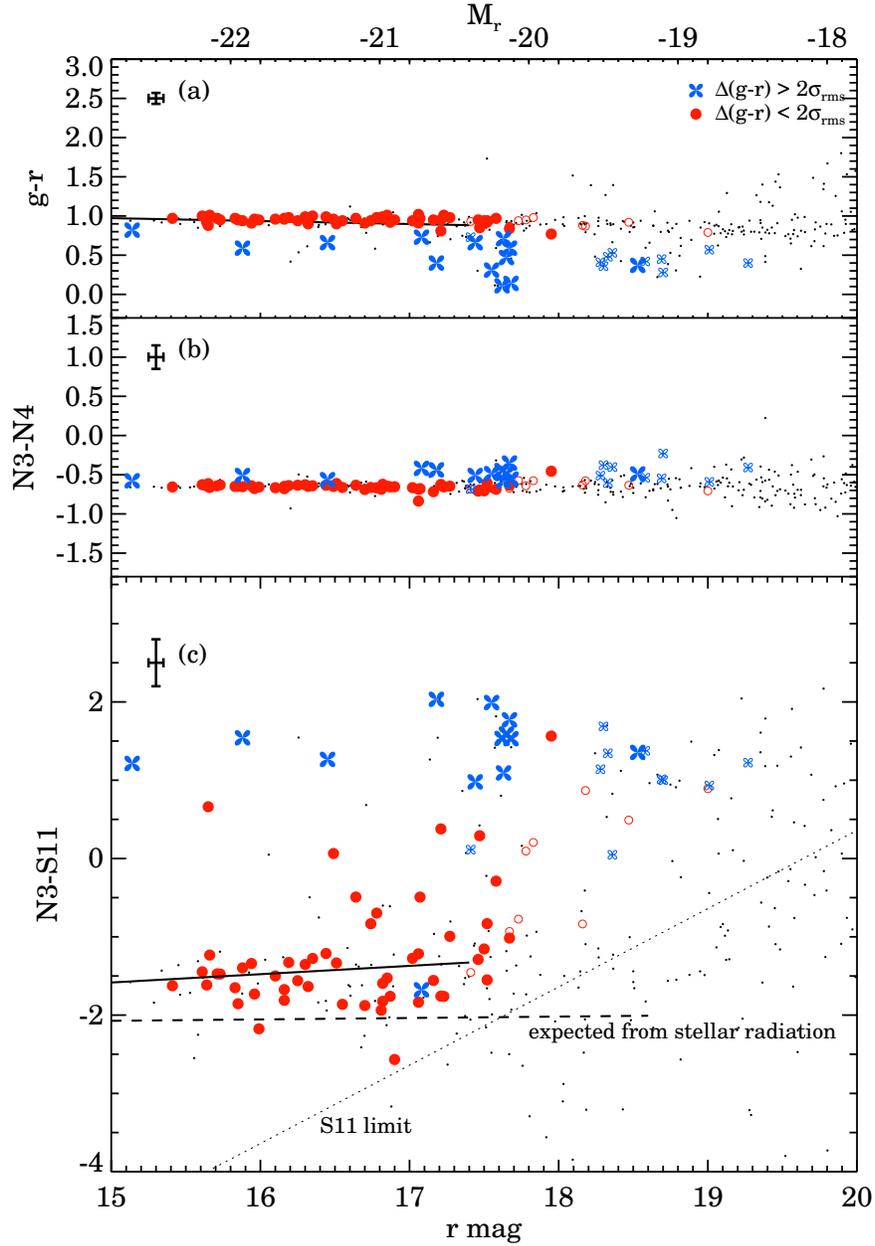}
 \caption{ \label{fig:cmr}
 Color-magnitude diagram of galaxies in the A2255 field 
 at different wavelengths. From top (a)
 to bottom (c), the $y$-axis of
 the color-magnitude diagram is ($g-r$), ($N3-N4$), and ($N3-S11$).
 The $x$-axis is identical to be $r$-magnitudes. 
  The colored points are galaxies detected in S11 
 ($f_{S11} > 80~\mu$Jy) : red circles indicate galaxies lying at
 optical red sequence, blue clovers indicate galaxies outside the
 red sequence. 
  Filled symbols are member galaxies with spectroscopic
 redshifts, and open symbols are galaxies with photometric redshifts.
  The small dots indicate all non-star objects in the observed field. 
 The lengths of $y$-axis is proportional to $y$-range for each panel.
 The error bar in top left of each panel indicates typical errors in 
 colors and magnitudes. The solid line in (a) and (c) is the 
 linearly-fitted color-magnitude relation (Eqn 1 and 2), and the  
 dashed line in (c) indicates the expected $N3-S11$ colors from 
 stellar radiation only. The dotted line in (c) is a $N3-S11$ 
 limit produced due to the S11 limit of $f_{S11} > 80~\mu$Jy.
 }
\epsscale{1.0}
\end{figure}

  It is well-known that cluster member galaxies show a clear
 red sequence in their color-magnitude relation defined by 
 elliptical galaxies, reflecting the similar age and/or metallicity
 of galaxies (e.g., Gladders et al. 1998). We investigate
 the color-magnitude relation of A2255 member galaxies at different
 wavelengths -- optical, NIR, and MIR. Figure \ref{fig:cmr} represents
 the color-magnitude diagram of A2255 member galaxies in $g-r$,
 $N3-N4$, and $N3-S11$, from top to bottom.

  In the optical and NIR (Figures \ref{fig:cmr}a and \ref{fig:cmr}b),
 there is a tight sequence of galaxies in the color-magnitude diagram.
  In the optical color-magnitude diagram, a tight sequence made by 
 bright, passively evolving member galaxies can be seen. The same
 sequence appears in 
 $r$ vs. $N3-N4$ color-magnitude relation, 
 but in a reversed way that the brighter galaxies
 having the bluer $N3-N4$ colors (Figure \ref{fig:cmr}b).
  We define the ``red sequence'' in the optical color-magnitude
 diagram to divide the member galaxies into two groups, 
 1) red-sequence galaxies 
 and 2) non red-sequence galaxies. 
  All the member galaxies with spectroscopic redshifts
 in the non red-sequence category show nebular emission lines 
 indicative of ongoing star formation, thus non red-sequence 
 galaxies are thought to be star-forming galaxies while red-sequence
 galaxies are not. 
  The color-magnitude relation is derived using a linear fit 
 to $r$ vs. $g-r$ relation for $r<17.5$ mag galaxies (see solid line in 
 Figure \ref{fig:cmr}a), by excluding the outliers iteratively based
 on the bi-weight estimator. The fitted color-magnitude relation is 
 as follows : 

  \begin{equation}
   g - r = ~ -0.037\times r~+ 1.53 
  \end{equation}

  The standard deviation of residuals to this fit is $\sigma_{rms}=0.08$ mag,
 indicating the tightness of optical red sequence. This value for scatter
 includes the photometric errors, and the slope in color-magnitude relation
 is consistent with that derived in other studies (e.g., Gallazzi et al. 2006).
  We consider galaxies to belong in the red sequence if they lie 
 within 
 $\Delta (g-r) < 2\sigma_{rms}$ from the red sequence, where 
 $\Delta (g-r)$ is an offset of color from the fitted relation.
  In Figure \ref{fig:cmr}, circles indicate optical red-sequence
 galaxies and clovers indicate non red-sequence galaxies.
  While the optical
 red sequence is produced by galaxies with similar ages and metallicities, 
 the same objects form a NIR ``blue'' sequence  
 since we are sampling the Rayleigh-Jeans tail of the black body
 radiation with the $N3-N4$ colors (e.g., Lacy et al. 2004; Stern
 et al. 2005).
 In NIR, the non red-sequence galaxies in the optical have 
 the redder $N3-N4$ colors than the $N3-N4$ blue sequence
 -- possibly due to the dust emission arising from star formation.

  In contrast, the tight sequence deteriorates in the color-magnitude
 relation in MIR, e.g., in the $r$ vs. $N3-S11$ color magnitude diagram
 (Figure \ref{fig:cmr}c). The spread in $N3-S11$ colors
 is much larger than those in optical or NIR,
 even if the outliers (clovers) from the optical red sequence are excluded.
  The linear relation derived from the $r$ vs. $N3-S11$ diagram (solid line in
 Figure \ref{fig:cmr}c) using the same method as the red sequence is : 

  \begin{equation}
  \begin{array}{l}
   N3-S11 = ~0.106\times r~- 3.18  
  \end{array}
  \end{equation}

  The scatter around this relation is $\sigma_{rms}=0.44$ mag, roughly six times 
 larger than that of in optical. The large scatter of $\sigma_{rms}=0.44$ mag
 in the MIR blue-sequence cannot be fully accounted for even if a maximum
 photometric error of $\sim0.3$ mag in $S11$ band is assumed.
 
  In addition to the large scatter, there are two interesting
 characteristics in the $N3-S11$ color-magnitude diagram. The first
 is that there is a weak ``blue'' sequence of galaxies, although
 the scatter is very large as described above. Contrary to the case
 of optical color-magnitude relation, the brighter galaxies have 
 the bluer MIR color. The second is that most of the galaxies 
 detected in the S11 band show the redder $N3-S11$ colors than
 the expected colors from stellar radiation alone
 ($N3-S11\sim-2.0$; dashed line in Figure \ref{fig:cmr}c).
  The value of $N3-S11 \sim -2.0$ is calculated using a model
 of Piovan, Tantalo, \& Chiosi (2003), where only
 stellar photospheric emission is considered -- 
 i.e. no dust continuum is taken into account. 
  Not only galaxies that are outliers of $N3-S11$ blue sequence, 
 but also galaxies in the blue sequence are considered to have 
 ``MIR-excess'' compared to the stellar continuum. The possible origin of
 the MIR-excess in these galaxies are the dust emission
 related to the star formation, AGN activity, and the circumstellar 
 dust shells around AGB stars (e.g., Ko et al. 2009), etc. 
  We investigate these possibilities by comparing MIR colors with
 galaxy model expectations in following section.

 \subsection{Color-Color Diagram} 

\begin{figure*}
\epsscale{1.1}
\plottwo{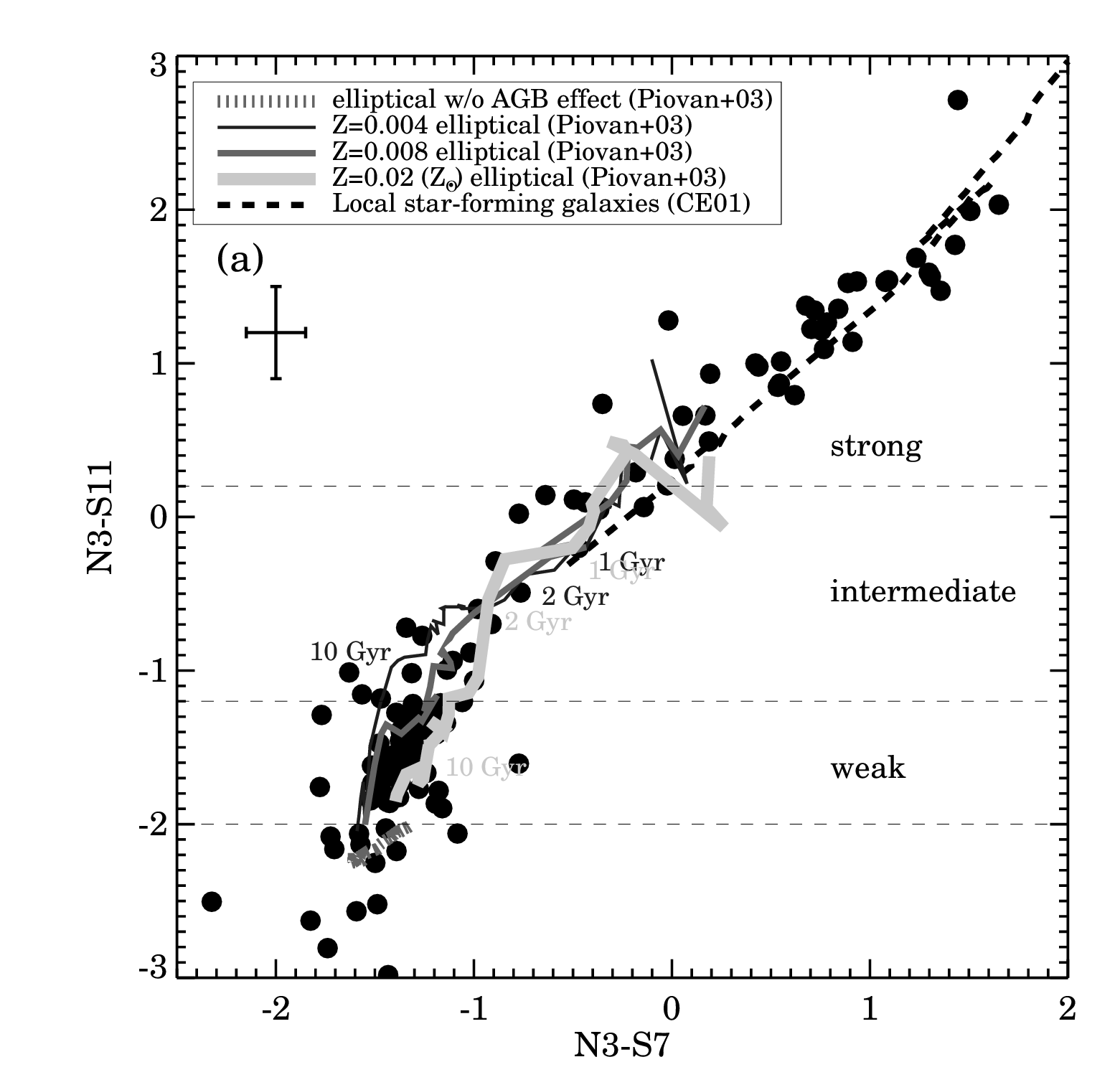}{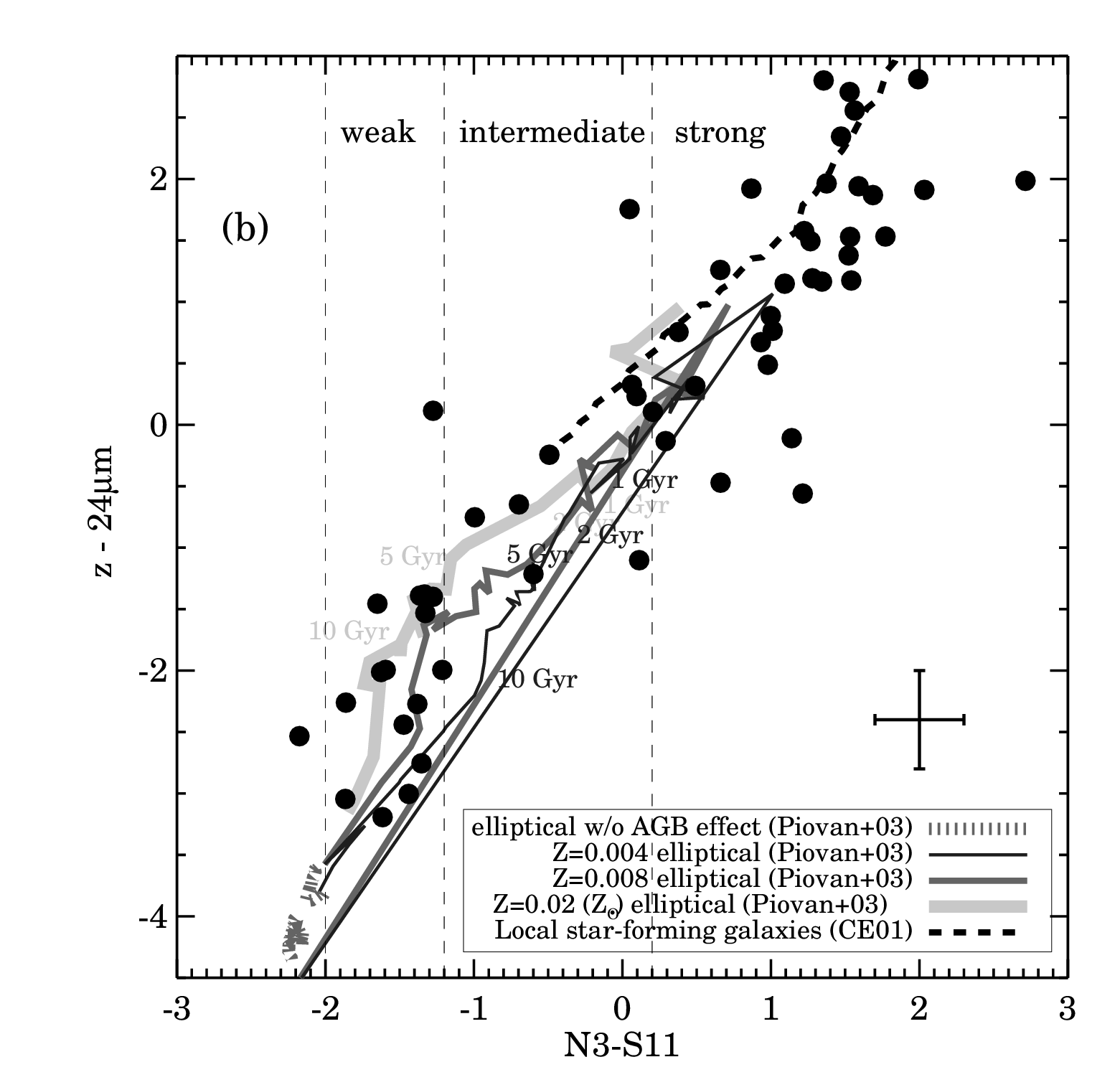}
 \caption{ \label{fig:mircolor}
 \textit{Left} : ($N3-S7$) vs. ($N3-S11$) color-color plot of A2255 member
 galaxies. Overplotted lines are the expected tracks of local star-forming
 galaxies (Chary \& Elbaz 2001; thick \textit{dashed} line), elliptical
 galaxies with different metallicities
 (Piovan, Tantalo, \& Chiosi 2003; \textit{solid} line).
 The metallicity increases from left ($Z=0.004$) to right ($Z=0.02$) in the
 $x$-axis, the age decreases as the ($N3-S11$) gets redder, and the total IR
 luminosity increases as ($N3-S7$) and ($N3-S11$) increases.
 There is an overlap between star-forming galaxy track and
 elliptical galaxies' tracks around (0,0).
 Filled circles are points from cluster member galaxies.
 The error bar indicates typical color errors produced by magnitude errors.
 \textit{Right} : The criteria for MIR-excess in terms of MIR-colors
 ($N3-S11$ and mag($z$)$-$mag(24\,$\mu$m)). As in Figure \ref{fig:mircolor}a,
 we use $N3-S11$ color as a measure of MIR-excess.
 Over the area where no 3\,$\mu$m or 11\,$\mu$m images are available, we use
 mag($z$)$-$mag(24\,$\mu$m) colors instead of $N3-S11$ colors. Filled circles 
 are points from cluster member galaxies with MIR-excess, while various
 lines are the expected relation from different models. 
 Again, the error bar indicates typical color errors.  
 }
\epsscale{1.0}
\end{figure*}

  In order to understand origins of MIR-excess in more detail, 
 we present a MIR color-color diagram of A2255 member galaxies in
 Figure \ref{fig:mircolor}a. The MIR colors used here are $N3-S7$ vs. $N3-S11$, 
 and the expected colors from various model galaxy templates are overplotted.
 The overplotted color tracks are calculated using models of elliptical 
 galaxies with different ages and metallicities which include the 
 dust emission from circumstellar dust around AGB stars 
 (Piovan, Tantalo, \& Chiosi 2003) or spectral energy distribution (SED)
 templates of local star-forming galaxies and IR luminous galaxies 
 (Chary \& Elbaz 2001).

  The plot shows that galaxies in the $N3-S11$ blue sequence
 ($\langle N3-S11 \rangle \sim-1.7$)
 are likely to be passively evolving galaxies with 
 stellar population age around 10\,Gyrs or larger.
 The reddest in both MIR colors
 ($N3-S11 >0.2$ mag and $N3-S7 > 0$ mag) are star-forming
 galaxies. Galaxies between two populations are dominated by 
 young, passively evolving galaxies with stellar ages between 1-10 Gyrs,
 yet these could also be galaxies with a small amount of star formation. 

  Based on the color tracks of model galaxy templates
 (Figure \ref{fig:mircolor}a), we classify MIR-excess galaxies into
 three classes according to the amount of MIR-excess. 
 The reddest galaxies, having $N3-S11 > 0.2$ mag are defined as 
 ``strong MIR-excess'', where the MIR color corresponds to that of
 an actively star-forming galaxy calculated using SED templates of
 Chary \& Elbaz (2001). 
 Galaxies at the $N3-S11$ blue sequence are 
 defined as ``weak MIR-excess'' galaxies ($N3-S11 < -1.2$ mag),
 which have MIR colors of passively evolving galaxies with old ages.
  The objects between the two populations are tagged as 
 ``intermediate MIR-excess'' galaxies whose MIR colors can have 
 multiple origins such as circumstellar dust emission from 
 intermediate age stars, residual star formation, and AGN activity.
 These three classifications based on the MIR color allows us to 
 investigate galaxy populations at different evolutionary stages. 
 The environmental dependence of galaxies is discussed in the 
 following sections based on these classifications.

  These MIR-excess terms are defined using $N3-S11$ colors and
 thus limited by N3 and S11 coverage (Figure \ref{fig:pointing})
 and depth.
  Therefore, we develop another criteria to define MIR-excess 
 using 24\,$\mu$m
 flux from \textit{Spitzer} MIPS, which covers a larger area
 than the S11 coverage.  Figure \ref{fig:mircolor}b shows 
 the correlation between $N3-S11$ and mag($z$)$-$mag(24\,$\mu$m) 
 for the observed galaxies, in addition to the expected model colors. 
  Two MIR colors, $N3-S11$ and mag($z$)$-$mag(24\,$\mu$m)  
 correlate reasonably well, and the locations of model tracks are 
 consistent with the case of $N3-S7$ vs. $N3-S11$. 
  Therefore we use mag($z$)$-$mag(24\,$\mu$m) to define MIR-excess
 galaxies in addition to $N3-S11$. The corresponding criteria
 for strong / intermediate / weak MIR-excess are, 
 mag($z$)$-$mag(24\,$\mu$m) $> -0.5$, $-2.0 <$ mag($z$)$-$mag(24\,$\mu$m) $<-0.5$,
 and $-3.5 <$ mag($z$)$-$mag(24\,$\mu$m) $<-2.0$ respectively.

  In Table \ref{tab:mirx_fraction}, we compare the number of
 cluster member galaxies with or without MIR-detection, and
 MIR-excess. The fractions of MIR-excess galaxies among cluster members 
 at the same $r$-band magnitudes are different for the AKARI S11 and 
 \textit{Spitzer} 24\,$\mu$m fields, e.g., 74\% for S11 and 37\% for
 24\,$\mu$m at $r<17.5$. This difference is due to 
 the shallower depth of the 24\,$\mu$m image compared to the S11 image. 
 As we showed in Figure \ref{fig:cmr}c, S11 flux limits or 24\,$\mu$m
 flux limits place limits on the MIR color that can be considered as 
 MIR-excess. With $5~\sigma$ flux limit of 80\,$\mu$Jy for S11 band, 
 weak MIR-excess galaxies are complete only at $r<17.5$, while
 intermediate and strong MIR-excess galaxies are complete down to
 $r\sim18.5$ mag. On the other hand at 24\,$\mu$m with flux limit of 
 250\,$\mu$Jy, the complete limit for weak MIR-excess galaxies is 
 $r<15$ mag, and the limit for intermediate MIR-excess galaxies 
 is $r<16.5$ mag. Sine our survey consists of fields covered with 
 different filters having different depths, we take this different 
 MIR-excess fraction into account when discussing the spatial distribution
 of MIR-excess galaxies (Section 5). 

  Table \ref{tab:mirx_fraction} as well as Figure \ref{fig:cmr}
 show that the fraction of MIR-excess galaxies is very high 
 ($>70$\% at $r<17.5$ mag, using S11), contrary to the general belief that 
 there are little dust and gas among cluster member galaxies.
  Many of these MIR-excess galaxies are on the optical red sequence 
 (weak and intermediate MIR-excess galaxies; see Figure \ref{fig:cmr}c),
 so that it is essential to include AGB circumstellar dust emission when 
 describe the SED of red cluster galaxies from optical through MIR
 (Bressan et al. 2007).
   This is also consistent with the trend in other galaxy cluster, 
 A2218, in which a significant MIR-excess is seen in fainter early-type
 member galaxies (Ko et al. 2009).

\subsection{Spectral Energy Distributions and Morphologies}

  For MIR-excess galaxies defined in Section 3.2,
 we calculate either
 total IR luminosity (i.e. IR-derived SFR) or
 stellar population age based on the SED fitting.
 The derivation of these quantities allow us to investigate
 star formation activities/quenching sequence and its relation to 
 the environment in A2255.
 The SED fitting follows the procedure described in
 Shim et al. (2007) except that
 we include the photometric data from optical 
 (SDSS $ugriz$) to MIR (AKARI IRC points, and \textit{Spitzer} MIPS
 24/70\,$\mu$m points if available).
  The template library we used is IR galaxy templates with
 different IR luminosities (Chary \& Elbaz 2001), and the early-type galaxy
 templates from Piovan, Tantalo, \& Chiosi (2003). 
  The IR galaxy templates are constructed through empirical interpolation
 between the observed local IR galaxies. As it is mentioned in Chary \&
 Elbaz (2001), the optical-NIR part of the SEDs are arbitrarily determined
 to match the IR vs. optical luminosity ratio,
 thus we used MIR (3-70\,$\mu$m) data points only
 to fit the IR SEDs. The early-type galaxy templates are
 constructed considering the effect of circumstellar dust around 
 AGB stars to the integrated spectrum in addition to the photospheric
 emission. The best-fit template is found using $\chi^2$ minimization
 between the model fluxes and the observed fluxes.

\begin{figure*}
  \epsscale{1.0} \plotone{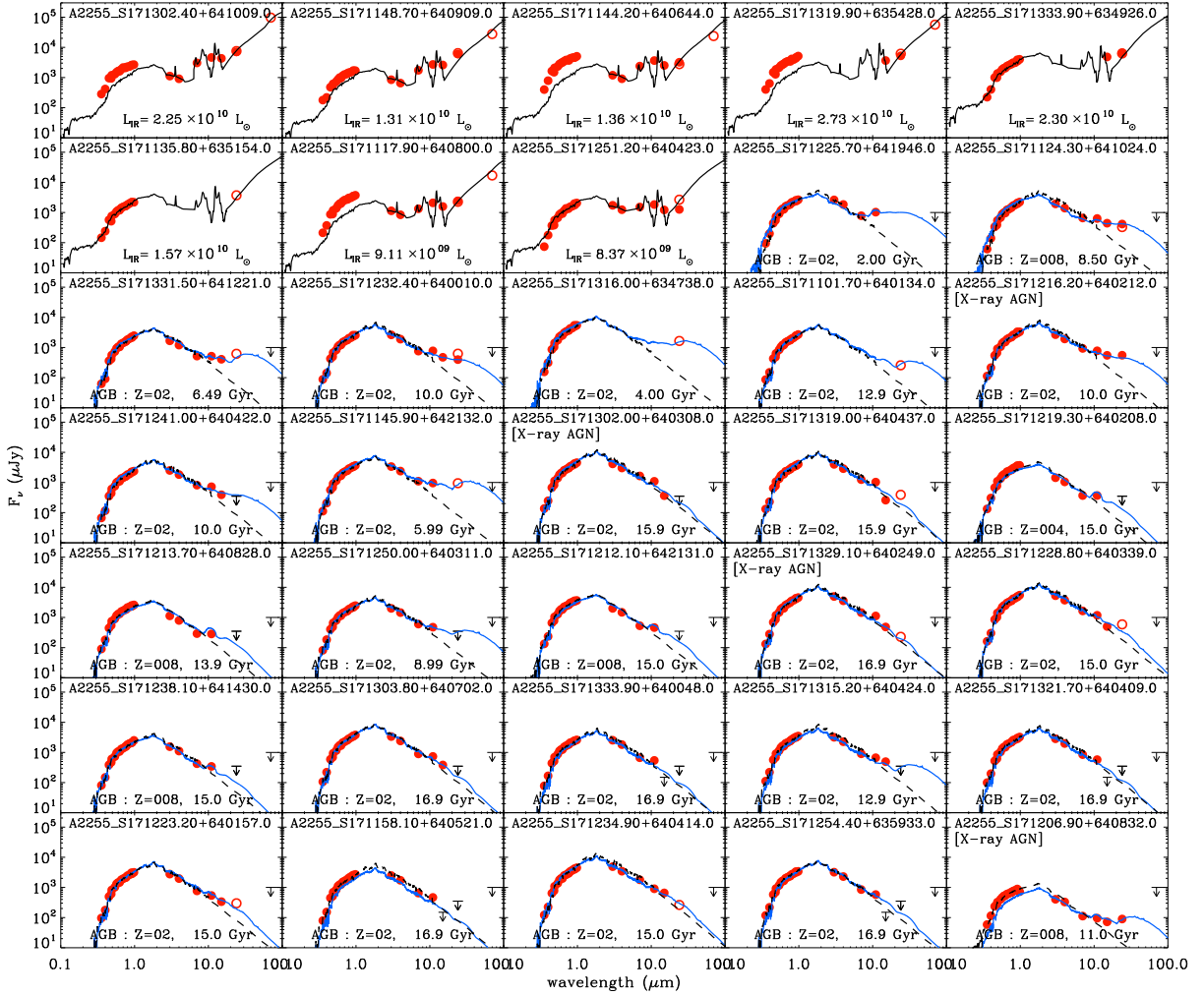}
  \caption{ \label{fig:seds}
  The SEDs of 34 ``optically'' bright ($r<16.5$ mag) galaxies in addition
  to one optically faint ($r\sim17.0$ mag) X-ray selected AGN.
  The morphologies of these galaxies are also
  presented in Figure \ref{fig:morph}, with the same order. 
  \textit{Filled} circles indicate optical and AKARI IRC photometry data points,
  while \textit{open} circles indicate MIPS 24\,$\mu$m and 70\,$\mu$m
  data points when available. In case we do not have detection despite
  the coverage, we mark flux upper limits with $5\sigma$ with arrows. 
  Overplotted lines represent the best-fit SEDs through the SED fitting
  using optical to MIR bands : the \textit{black solid} lines indicate 
  IR galaxy templates of Chary \& Elbaz (2001), \textit{blue solid}
  lines indicate the early-type galaxy templates considering the effect
  of AGB dust (Piovan, Tantalo, \& Chiosi 2003), and \textit{black dashed}
  lines indicate early-type galaxy templates without AGB dust effect 
  (Piovan, Tantalo, \& Chiosi 2003). Note that for star-forming galaxies,
  we only used MIR (3-70\,$\mu$m) photometry data points in SED fitting
  since the optical-NIR part of the SEDs was arbitrarily defined in 
  Chary \& Elbaz (2001).
  }
\end{figure*}

\begin{figure*}
  \epsscale{0.8} \plotone{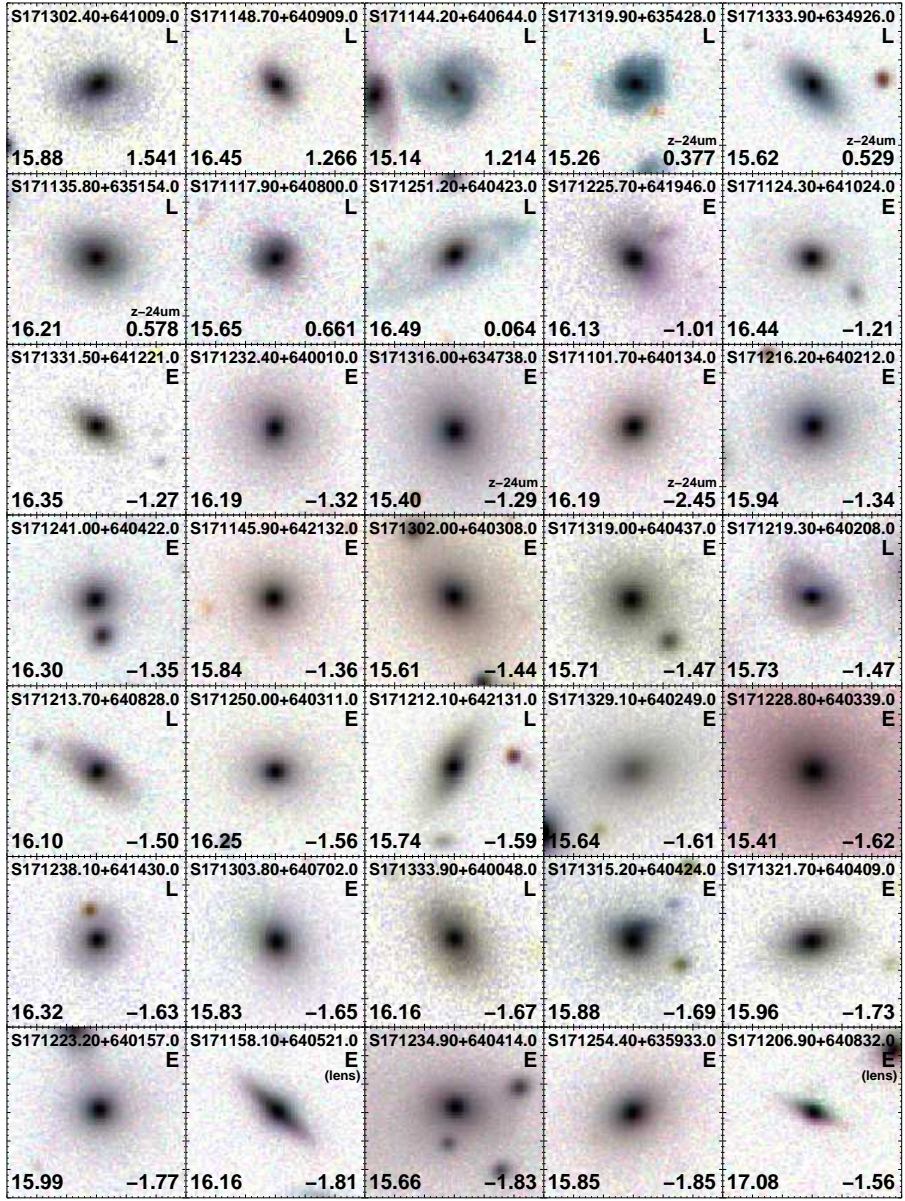}
  \caption{ \label{fig:morph} The morphologies of 34 objects with
  $r<16.5$ mag and one X-ray selected AGN with $r=17.08$ mag.
  The stamp images are color-composites of SDSS $g$, $r$, and $i$-band
  (inverse) images, with the size of $30\arcsec\times30\arcsec$.
  It corresponds to $45~\mbox{kpc}\times45~\mbox{kpc}$ in physical scale.
   Indicated in the postage stamp images are the object id, 
  morphological type (E for early-type, and L for late-type), 
  $N3-S11$ color,
  and the $r$-band magnitudes (from top to clockwise direction). 
  For those without $S11$ photometry due to the area coverage, 
  mag($z$)$-$mag(24\,$\mu$m) colors are given instead. 
   The objects are sorted in the order of decreasing MIR-excess
  (i.e., decreasing $N3-S11$ colors), except the last AGN object.
  }
\end{figure*}

  Figure \ref{fig:seds} shows the examples of the SED fitting for 34 relatively
 bright galaxies in optical ($r<16.5$ mag, i.e. $M_r < -21.0$ mag).
 These SED panels are arranged in a descending order of $N3-S11$ values. 
 Since these galaxies are sufficiently bright and large, we also present 
 their color-composite stamp images in Figure \ref{fig:morph} as a guide
 to their morphologies. 
 The images are made using SDSS $i$-band images by overlaying $g$-band 
 as blue color. The size of each cutout is 
 $30\arcsec \times 30\arcsec$, and the images are displayed in logarithmic 
 scale. 
  The objects in Figure \ref{fig:morph} are aligned with the same order 
 as Figure \ref{fig:seds}. At $r>16.5$ mag, it becomes  
 difficult to investigate the morphologies using the SDSS images
 unless they are sufficiently extended.
  The last object in Figure \ref{fig:seds} and \ref{fig:morph} does not
 fall in the magnitude cut for optically bright objects,
 yet the object is included as an example of member galaxies 
 with known AGNs showing MIR-excess.

  The MIR ($N3-S11$) colors are marked in the lower right of 
 each cutout image in Figure \ref{fig:morph}.
  The first seven galaxies are strong MIR-excess galaxies, and the 
 eighth galaxy is an intermediate MIR-excess galaxy. These eight
 galaxies are best-fitted with IR galaxy templates (Chary \& Elbaz 2001)
 showing late-type morphology or disky features.
  All the remaining galaxies are weak MIR-excess galaxies, which are
 best-fitted by early-type galaxy templates of various ages and 
 metallicities. Although there are two free parameters -- metallicity
 and age -- for early-type galaxy templates, we only use age as a 
 meaningful parameter from fitting since the choice of metallicity is
 very limited ($Z=$ 0.004, 0.008, and 0.02). 
  Again, while the age itself is a model-dependent parameter
 (thus there exists ``unphysical'' age which is larger than the age of
 universe), we see that the amount of MIR-excess is mainly affected by
 the age of a galaxy.

\subsection{The Nature of Intermediate MIR-Excess Galaxies}

  In Section 3.1 (Figure \ref{fig:cmr}), we show that there is a
 large number of galaxies that show MIR-excess while lying on the
 tight optical red sequence at the same time. They fall on to the
 ``intermediate'' ($-1.2<N3-S11<0.2$) or ``weak'' ($-2.0<N3-S11<-1.2$)
 MIR-excess categories (Figure \ref{fig:mircolor}).
  To check the origin of the weak/intermediate MIR-excess, and
 provide a physical meaning of the intermediate MIR-excess galaxies,
 we investigate various properties of this population. 

  First, we checked the optical spectra of the intermediate MIR-excess
 galaxies with $r$-band magnitude brighter than 17.77mag. We find 
 no sign of emission lines, which indicates that the MIR-excess of
 most of these galaxies is not due to the star formation activity
 (at least within the fiber aperture). 

\begin{figure}
 \epsscale{1.0} \plotone{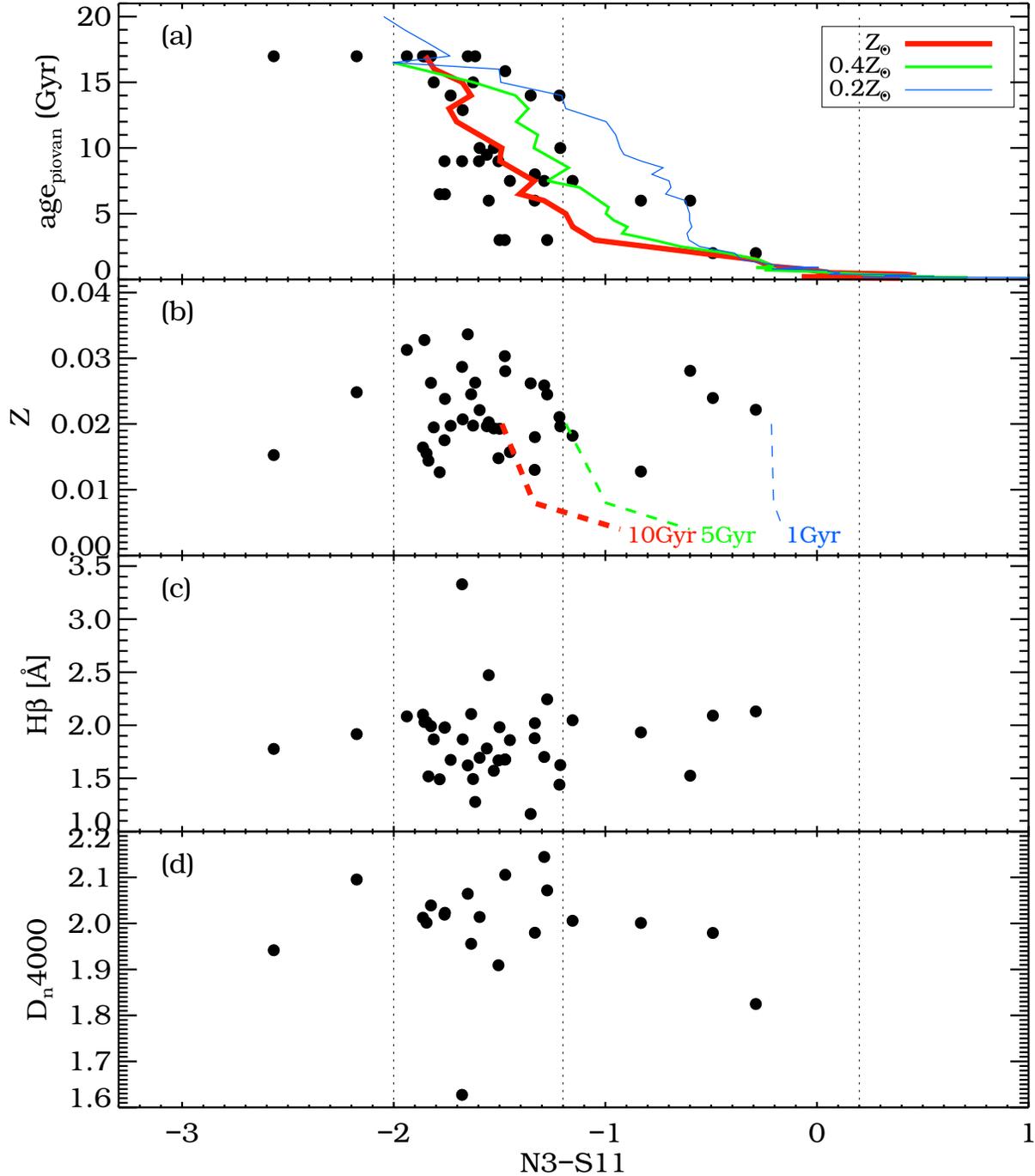}
 \caption{\label{fig:mirx}
  The relation between MIR excess ($x$-axis, $N3-S11$) and (a) age
 derived from SED fitting, (b) metallicity measured in the SDSS spectra
 (Gallazzi et al. 2005), (c) H$\beta$ line indices measured using SDSS spectra
 (SDSS-MPA catalog; Section 2.4), and (d) D$_n$(4000) measured using SDSS
 spectra (Kauffmann et al. 2003).
 \textit{Filled} circles represent member galaxies with metallicity, the
 H$\beta$, and/or D$_n$(4000) measurements. The number of objects with
 metallicity and H$\beta$ is 40 (panel a, b, and c), while there are
 20 galaxies with D$_n$(4000) measurements in addition (panel d). 
 Strong MIR-excess ($N3-S11>0.2$) 
 galaxies are not included in these plots since their MIR-emission 
 mechanism is different from that of weak/intermediate 
 MIR-excess galaxies. 
 The dotted vertical lines indicate the
 criteria for dividing weak/intermediate/strong MIR-excess, 
 i.e., $N3-S11=-2.0, -1.2,$ and $0.2$.
 }
 \epsscale{1.0}
\end{figure}

  We also examined the relation between MIR-excess and other 
 spectral properties of red sequence galaxies (including weak MIR-excess
 galaxies). 
  The top panel of Figure \ref{fig:mirx} shows the $N3-S11$ colors vs.
 the luminosity-weighted mean stellar age derived from the SED fitting
 (see Section 3.3). The second panel from the top illustrates the
 $N3-S11$ colors vs. the metallicities measured from the SDSS spectra
 (Gallazzi et al. 2005), of next is the $N3-S11$ colors vs. the H$\beta$
 absorption line equivalent widths (Gallazzi et al. 2005), and the
 final panel shows the $N3-S11$ colors vs. D$_n$4000 (Kauffmann et al. 2003).
  The SED-fitted ages are derived by fixing metallicity to the solar value,
 which should be a good approximation judging from Figure \ref{fig:mirx}b.
 Due to the limitation in modeling, there are ``unphysical'' ages
 that are larger than the age of the universe ($\sim17$ Gyr).  The absolute
 age of the SEDs should not be taken too seriously, as these models
 are not meant to provide absolute ages (Piovan, Tantalo, \& Chiosi
 2003). These model fit parameters are meant to provide the relative
 age scales represented by different SED shapes.
  We find that the ages of the intermediate/weak MIR-excess galaxies
 derived by the SED fitting correlate with the $N3-S11$ colors 
 (Figure \ref{fig:mirx}a),
 with the Spearman's rank correlation coefficient of $-0.71$.
 Considering the degree of freedom at 38 (N-2, the number of points used
 is 40), the correlation is reliable by more than 99.9\% (Zar 1972). 
  The correlation is already expected by the model color tracks overplotted
 in Figure \ref{fig:mirx}a. On the other hand, there is little
 correlation between the metallicity and the $N3-S11$ color
 (Figure \ref{fig:mirx}b; $r_s = 0.008$, consistent with null hypothesis).  
  Since the $N3-S11$ colors
 are more sensitive to age than metallicity, the MIR-excess can be
 used to break the old age-metallicity degeneracies (e.g., Ko et al.
 2009).
  Figure \ref{fig:mirx}c and \ref{fig:mirx}d show the relation
 between MIR-excess and other well-known age indicators, the equivalent
 widths of Balmer absorption line (H$\beta$) and D$_n$(4000),
 a measure of the strength of the $4000\mbox{\AA}$ break 
 (Kauffmann et al. 2003). 
 The Spearman's rank correlation
 coefficient in this case is $r_s=-0.07$ and $-0.14$, i.e., significance
 level (of the rejection of null hypothesis) greater than 50\% (degree of freedom at 38)
 and 40\% (degree of freedom at 18) respectively. 
 Thus, the correlation between the MIR color and H$\beta$, D$_n$(4000)
 is relatively weak. 

\begin{figure}
 \epsscale{1.0} \plotone{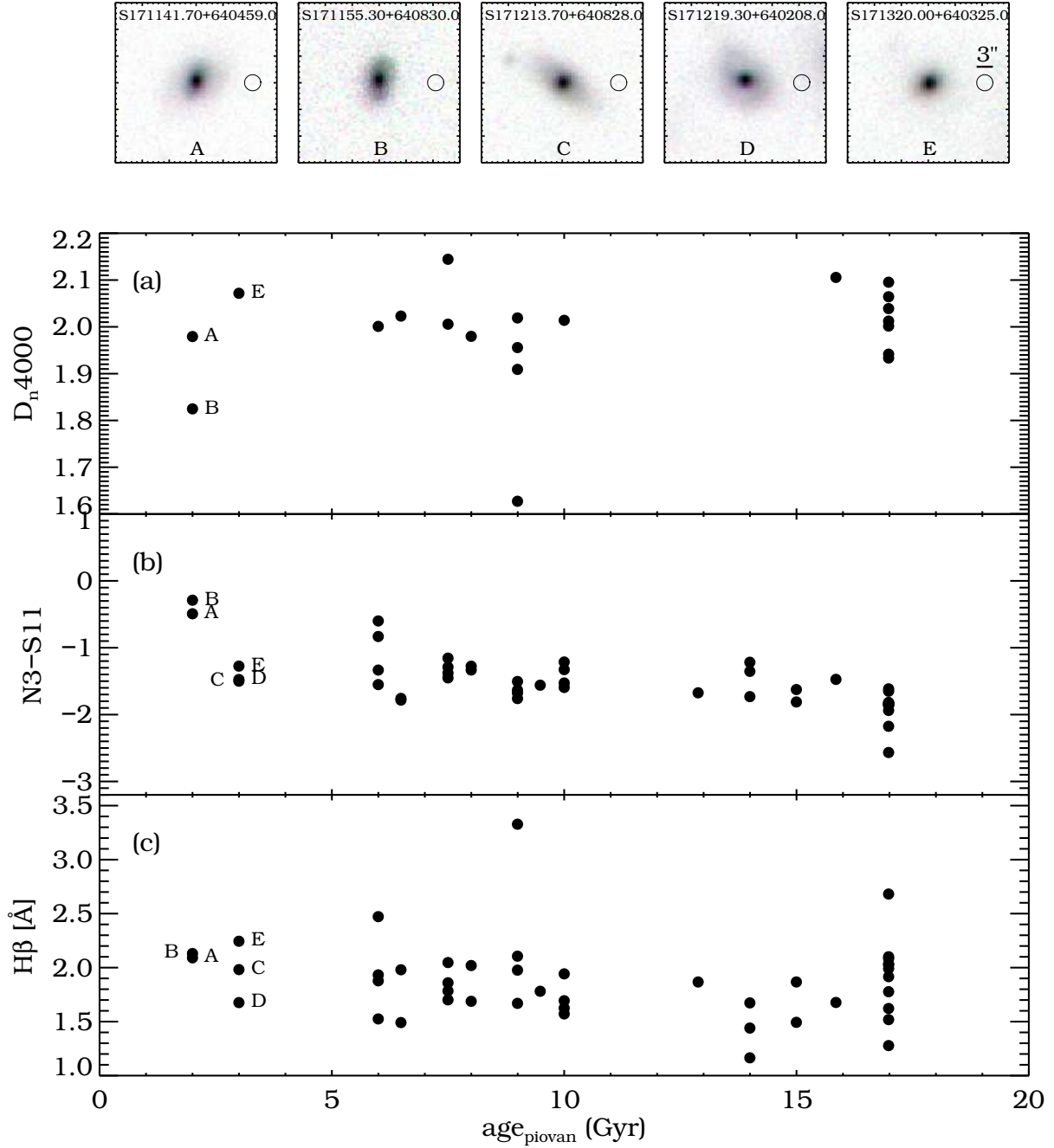}
 \caption{\label{fig:ageindx}
  The relation between age derived SED fitting ($x$-axis) and
 (a) D$_n$(4000), (b) $N3-S11$, and (c) H$\beta$ indices of 
 intermediate / weak MIR-excess galaxies. 
  The points plotted are the same as those in Figure \ref{fig:mirx}.
 The different
 stellar age indicators (D$_n$4000, $N3-S11$, and H$\beta$)
 correlate with the stellar age but with large scatters. The SDSS $r$-band
 images of the youngest ($<5$\,Gyr) galaxies are illustrated above
 to show their morphologies. The small circle in each image 
 represent the SDSS fiber size, $3\arcsec$.  }
 \epsscale{1.0}
\end{figure}

\begin{figure}
\plotone{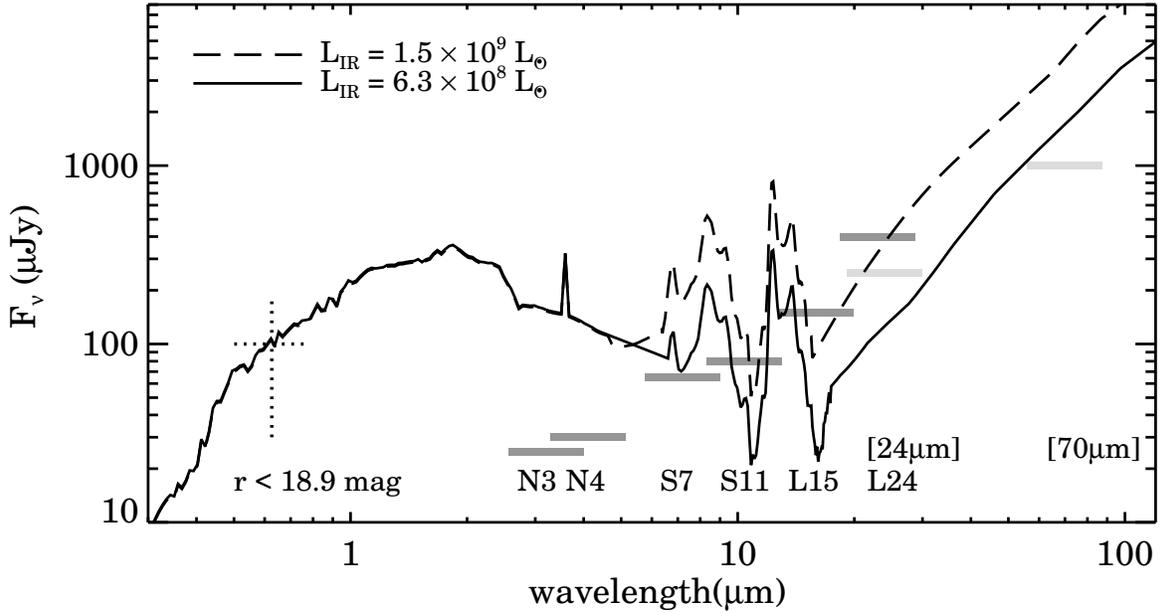}
 \caption{ \label{fig:irlimit}
   The comparison between star-forming galaxy templates with different
  IR luminosity and the flux limits in each filter. The overplotted lines
  are star-forming galaxy templates with $L_{IR} = 6.3\times10^8~L_{\odot}$
  (\textit{dashed} line) and 
  $L_{IR} = 1.5\times10^9~L_{\odot}$ (\textit{solid} line). The limits 
  (shaded region) indicate $5\sigma$ flux limits over $2\times$FWHM
  diameter in all filter bands (see Table \ref{tab:obs_summary}).
   When optical magnitude limit for complete redshift identification, 
  $r < 18.9$ mag (see Section 2.2) is applied, the limits in the derived
  IR luminosity varies between 
  $L_{IR} > (0.6-1.5)\times10^9~L_{\odot}$; On average, the minimum SFR
  we derive through SED fitting is $>0.1~M_{\odot} \mbox{yr}^{-1}$
  according to the IR luminosity limit of $L_{IR} > (0.6-1.5)\times10^9~L_{\odot}$. 
  To be conservative, we treat that our MIR data is complete above 
  $L_{IR} > 1.5\times10^9~L_{\odot}$.
  }
\end{figure}

  Figure \ref{fig:ageindx} illustrates the discrepancy between the
 different age indicators more clearly. The correlation coefficients
 between the SED-fitted age and age indicators are $r_s = 0.14, -0.70,$
 and $-0.18$ for D$_n$(4000), $N3-S11$, and H$\beta$ equivalent width
 respectively. These represent significance levels of smaller than 40\%, 
 larger than 99\%, and smaller than 70\% respectively. Only $N3-S11$ 
 shows reliable correlation with the stellar age derived through 
 SED fitting. This implies that the estimation of mean stellar age 
 is not an easy task.
  Each age indicator has its pros and cons : H$\beta$
 is known to be less sensitive to metallicity compared to a simple
 color index such as $B-V$ as viewed from the stellar population
 synthesis modelling, yet it is also easily affected by emission line
 component. D$_n$(4000) is known to be a good tracer for young stellar
 age ($<1$\,Gyr, D$_n(4000) < 1.5$; Balogh et al. 1999), yet highly
 affected by metallicity for older age (Kauffmann et al. 2003). 
  Moreover, both H$\beta$ and
 D$_n$(4000) is limited by the finite fiber size used for taking
 the SDSS spectra ($3\arcsec$ diameter) that only samples the central
 part of a galaxy if it is extended. We show morphologies of five galaxies
 with the mean stellar ages from the SED fitting younger
 than 5\,Gyr (Figure \ref{fig:ageindx}).
 Most of the galaxies have outer disks and late-type morphology,
 and finite SDSS fiber size sample only the central regions as indicated
 as $3\arcsec$ bar in Figure \ref{fig:ageindx}.
  Therefore, we address that MIR-excess is a relatively good age
 indicator that is free of finite spectroscopic aperture, and 
 age-metallicity degeneracy.

  As the morphologies of 5 young galaxies ($<5$\,Gyr) suggests
 (Figure \ref{fig:ageindx}), the fraction of late-type morphologies is
 large for ``intermediate'' MIR-excess galaxies ($-1.2<N3-S11<0.2$).
 We examine the morphologies of A2255 member galaxies at the bright
 end ($r<16.5$\,mag; Figure \ref{fig:morph}), since the visual
 morphological classification becomes difficult at the fainter 
 magnitudes. While 7 out of 7 strong MIR-excess galaxies
 display late-type galaxy morphology, only 6 out of 25 (24\%)
 weak MIR-excess galaxies show late-type morphology which can be
 easily understood as a consequence of the well-known correlation
 between color and morphology of galaxies. In the case of the intermediate
 MIR-excess galaxies (S171251.20+640423.0 and S171225.70+641946.0 
 in Figure \ref{fig:morph} and galaxies in Figure \ref{fig:ageindx}), 
 they show late-type morphology such as disks despite of being at the
 optical red sequence. 
  The natural explanation of the colors and the morphologies of
 the intermediate MIR-excess galaxies is that these galaxies are
 late-type galaxies with small amount of star formation, very
 possibly in the process of quenching of star formation. 
  Studying E$+$A galaxies (i.e., characterized with old stellar population 
 and strong Balmer absorption) would be another way to see the
 truncation of star formation, yet we did not find any E$+$A galaxies
 using a criteria of EW(H$\delta$)$>5~\mbox{\AA}$ (Goto 2007, 
 in addition to the criteria for no detectable [OII] and H$\alpha$
 emission lines) among the member galaxies with the spectroscopic
 information. The intermediate MIR-excess galaxies have on average 
 EW(H$\delta$) of $\sim1~\mbox{\AA}$, with a typical measurement
 uncertainty of $0.5~\mbox{\AA}$.
  Only $\sim50$\% of the intermediate MIR-excess galaxies are 
 spectroscopically observed, thus it is not clear whether the remaining
 $\sim50$\% would be classified as E$+$A galaxies or not. 
  This result demonstrates an advantage of using MIR-excess in 
 studying the transition population over optical spectra. 
  The ``weakened'' star formation activity in intermediate 
 MIR-excess galaxies, regardless of its origin (weak star formation
 or the AGB-dust), is naturally explained by the idea that 
 these galaxies are in the 
 process of transformation from star-forming galaxies (strong MIR-excess
 galaxies) to quiescent galaxies (weak MIR-excess galaxies) at optical
 red sequence. The relation with these transformation and the cluster
 environment will be described in more detail in Section 5.3.

 \subsection{Contribution of Active Galactic Nuclei to MIR-Excess Galaxies Sample}

  Active Galactic Nuclei cause MIR-excess in galaxies (Quillen et al. 1999)
 as well as ongoing star formation, thus alter the derived star formation
 rates of galaxies. The different diagnostics for AGN, such as infrared 
 color and optical line ratio, produce a significantly
 different result for demarcating AGNs and star-forming galaxies among 
 MIR-excess galaxies (Brand et al. 2009). Therefore, we do not force to 
 differentiate AGNs from our MIR-excess galaxies in this paper. Instead, 
 we briefly discuss how many AGNs are included in MIR-excess galaxies
 and how they affect our analysis and conclusion,
 using the previously known AGNs detected in either X-ray or radio, classified
 as AGNs using optical line diagnostics (Miller \& Owen 2003;
 Davis, Miller, \& Mushotzky 2003). 

  In the AKARI/\textit{Spitzer} field of view, there lies 13 previously
 known AGNs. Among the 13 AGNs, 10 are detected in 
 in either 11\,$\mu$m (S11) or 24\,$\mu$m (L24/MIPS 24\,$\mu$m) images,
 and 3 are not. Ten AGNs detected in 11\,$\mu$m or 24\,$\mu$m are radio
 source with 1.4\,GHz flux larger than 0.35\,mJy, while three are Seyferts 
 and seven are objects with old stellar population either with weak 
 [NII] or [SII] emission lines, or without any signs of emission lines 
 (Miller \& Owen 2003).
  By number, AGN contributes 3\% of the star-forming galaxies, 
 and 15\% of the weak MIR-excess galaxies. No AGN cross-identification is 
 found in intermediate MIR-excess galaxies. The contribution from three 
 Seyfert galaxies to the total SFR is less than 5\,$M_{\odot}$ yr$^{-1}$, i.e.
 less than 4\% of $\sum_{>0.25~M_\odot yr^{-1}} SFR $
 (see Section 4.2 for details). With this little contribution, we 
 conclude that our main analysis and conclusion is not affected
 by the presence of AGNs among MIR-excess galaxies
 especially in the sense that we mainly discuss star-forming galaxies
 and intermediate MIR-excess galaxies in A2255.

\section{Star Formation Rates}
 
  As we have shown in the previous section, the strong MIR-excess
 galaxies ($N3-S11>0.2$) are galaxies with ongoing star formation.
 In this section, we discuss how much star formation is hidden in IR
 (e.g., Bai et al. 2007), 
 and compare the total SFR of A2255 with those of other galaxy clusters 
 too see if there is anything special in the star formation activity
 of A2255. Only galaxies with strong MIR-excess are considered here 
 for the SFR study, since SFRs in other classes of MIR-excess galaxies 
 are minimal, less than 0.1\,$M_{\odot}$ yr$^{-1}$.

  \subsection{Infrared Star Formation Rate}

  We calculate SFRs of member galaxies from the derived IR luminosity 
 through the SED fitting (Section 3.3, Figure \ref{fig:seds}), 
 and compare the SFR with those from other wavelengths. 
  The total IR luminosities of A2255 member galaxies range from 
 $1.0\times10^8$ to $3.2\times10^{10}~L_{\odot}$,
 while the median value is $\langle L_{IR}\rangle\simeq2.8\times10^9~L_{\odot}$.
 These values correspond to
 $\sim0.1~M_{\odot} \mbox{yr}^{-1} < SFR (IR) \lesssim 7.0~M_{\odot} \mbox{yr}^{-1}$
 when converted using Kennicutt (1998) relation between $L_{IR}$ and SFR. 
  However, considering the flux limits in MIR and cluster member 
 completeness limit in optical, the conservative limit in $L_{IR}$
 in our survey is $L_{IR}>1.5\times10^9~L_{\odot}$ (Figure \ref{fig:irlimit}). 
 We do not find any Luminous IR Galaxy (LIRG; $L_{IR} > 10^{11}~L_{\odot}$)
 candidates in A2255.

\begin{figure*}
\plotone{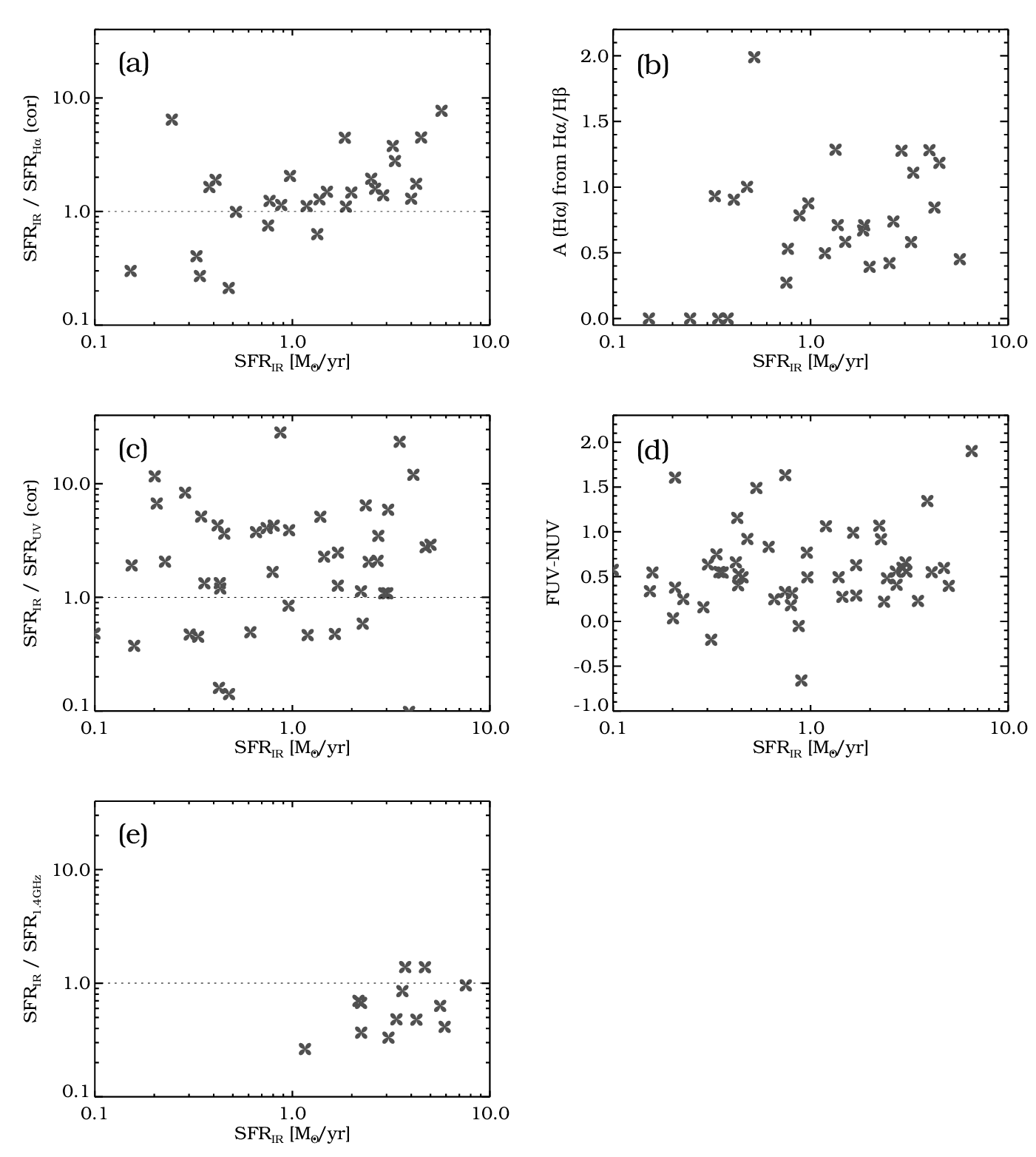}
 \caption{ \label{fig:sfr_extinct}
 The ratio between SFR$_{IR}$ and SFR from different indicators
 (a: SFR$_{IR}$/SFR$_{H\alpha}$, c: SFR$_{IR}$/SFR$_{UV}$,
 e: SFR$_{IR}$/SFR$_{1.4GHz}$) in addition
 to the dust attenuation measures in $H\alpha$ (b) and $UV$ (d).
 In (a) and (b), SFR$_{H\alpha}$ and SFR$_{UV}$ are extinction corrected
 using Balmer decrement and UV slope $\beta$ respectively.
 SFR$_{H\alpha}$ is derived using the H$\alpha$ line flux measured in SDSS
 optical spectra. The mean extinction from Balmer line ratio is
 $\langle\mbox{A}(H\alpha)\rangle \sim 0.78$ mag, suggesting
 a factor of $\sim2$ extinction in SFR$_{H\alpha}$.
 $(FUV-NUV)$ color is another measure of dust extinction that reflects
 UV slope $\beta$.
 }
\end{figure*}

  We compare the ``IR-derived SFR'' with SFRs derived using other 
 star formation indicators (Figure \ref{fig:sfr_extinct}) at different
 wavelengths. Figure \ref{fig:sfr_extinct}a shows the comparison between
 $SFR_{IR}$ and $SFR_{H\alpha} (cor.)$, i.e. IR-derived SFR and the
 H$\alpha$-derived SFR corrected for dust extinction
 (``cor.'' means the luminosity is corrected for internal dust extinction).
  In order to derive $SFR_{H\alpha} (cor.)$,
 we first derive $SFR_{H\alpha} (uncor.)$, SFR based on H$\alpha$ flux
 not corrected for the internal dust extinction, 
 using H$\alpha$ fluxes from the
 SDSS-MPA catalog (see Section 2.4) and the H$\alpha$ SFR conversion formula
 of Kennicutt (1998). Since the SDSS spectra are obtained using 
 fibers with a fixed size of $3\arcsec$ diameter, we adopt the aperture correction
 for each object from Brinchmann et al.(2004), i.e.
 $f_{total}/f_{fiber}$\footnote{Note that
 Brinchmann et al.(2004) also provide the spectroscopic SFRs
 by spectral line fitting; however we find that their SFRs significantly 
 depend on the optical colors of galaxies, reflecting the past star
 formation history of galaxies. Since we focus on the current ongoing
 SFR in A2255 members, we just use the H$\alpha$ line to derive optical 
 SFR.}.
  The ``extinction'' in H$\alpha$ flux -- A($H\alpha$) -- 
 is estimated from the Balmer decrement ($H\alpha/H\beta$ flux ratio),
 by adopting a case B recombination at $T=10,000K$ (Osterbrock 1989)
 and the assumption of the Calzetti (2000) extinction law for starburst galaxies.  
  The median amount of extinction in H$\alpha$ is 
 $\langle A(H\alpha) \rangle = 0.78$ mag (Figure \ref{fig:sfr_extinct}b). 
 There is a weak correlation between SFR and $A(H\alpha)$, i.e.
 galaxies with larger SFR having larger A($H\alpha$). 

    We also compare the UV-derived SFR with the IR-derived SFR 
 (Figure \ref{fig:sfr_extinct}c), using the GALEX NUV flux as a measure of
 SFR$_{UV}$.
  As in the case of H$\alpha$, we derive the extinction correction in 
 UV wavelengths from the UV slope $\beta$ inferred by $FUV-NUV$ colors
 (Meurer, Heckman, \& Calzetti 1999). Star-forming galaxies
 in A2255 show relatively blue $FUV-NUV$ colors of
 $\langle FUV-NUV \rangle = 0.55$ mag (Figure \ref{fig:sfr_extinct}d).
  The mean extinction correction derived is
 $\langle A(UV) \rangle \sim1.2$ mag. When this correction is applied
 (SFR$_{IR}$ vs. SFR$_{UV} (cor.)$; Figure \ref{fig:sfr_extinct}c), 
 the two SFRs are comparable with each other. 
  Both the H$\alpha$- and the UV-derived SFRs agree with the IR-derived 
 SFR when corrected properly for the internal dust extinction, albeit
 with a large scatter ($> \times2$).

  Finally in Figure \ref{fig:sfr_extinct}e, we present the comparison
 between SFR$_{IR}$ and SFR$_{radio}$ using 1.4\,GHz flux in
 Miller \& Owen (2003).
 The conversion formula from 1.4GHz flux to SFR$_{radio}$ is 
 adopted from Bell (2003). Although the number of the matched
 member galaxies is small, SFRs from the two indicators
 differ by about a factor of two, with the SFR$_{IR}$ being systematically
 larger than SFR$_{radio}$. 

  Overall, we find a conventional extinction correction to the dust 
 extinction works roughly well for estimating SFRs in individual 
 galaxies (e.g., Choi et al. 2006), but such correlations accompany 
 large scatters. Therefore, we consider the usage of the IR data is 
 a robust way to derive the star formation compared to UV and the 
 optical data.

  \subsection{Total Star Formation Rate of A2255}

  In this section, we derive the global properties of star formation
 activity in A2255, in order to examine if the merging activity 
 enhanced the star formation activity of A2255 in particular. This is
 done by deriving the ``total'' SFR in A2255 by constructing the IR
 luminosity function and comparing the total SFR of A2255 with other 
 galaxy clusters. The construction of the luminosity function 
 is done using the following equation 

\begin{equation}
  \phi (log L_i) = \frac{1}{A} \frac{n}{\Delta(log L)}
\end{equation}

\noindent while $A$ is the surveyed area,
 $n$ is the number of galaxies whose luminosity falls 
 within the $i$-th bin, and $\Delta(log L)$ is a luminosity bin size.
 We used the survey coverage of $\sim2000$\,arcmin$^2$, i.e.
 $\sim16$\,Mpc$^2$ in physical scale. 

\begin{figure*}
  \epsscale{1.2}
  \plottwo{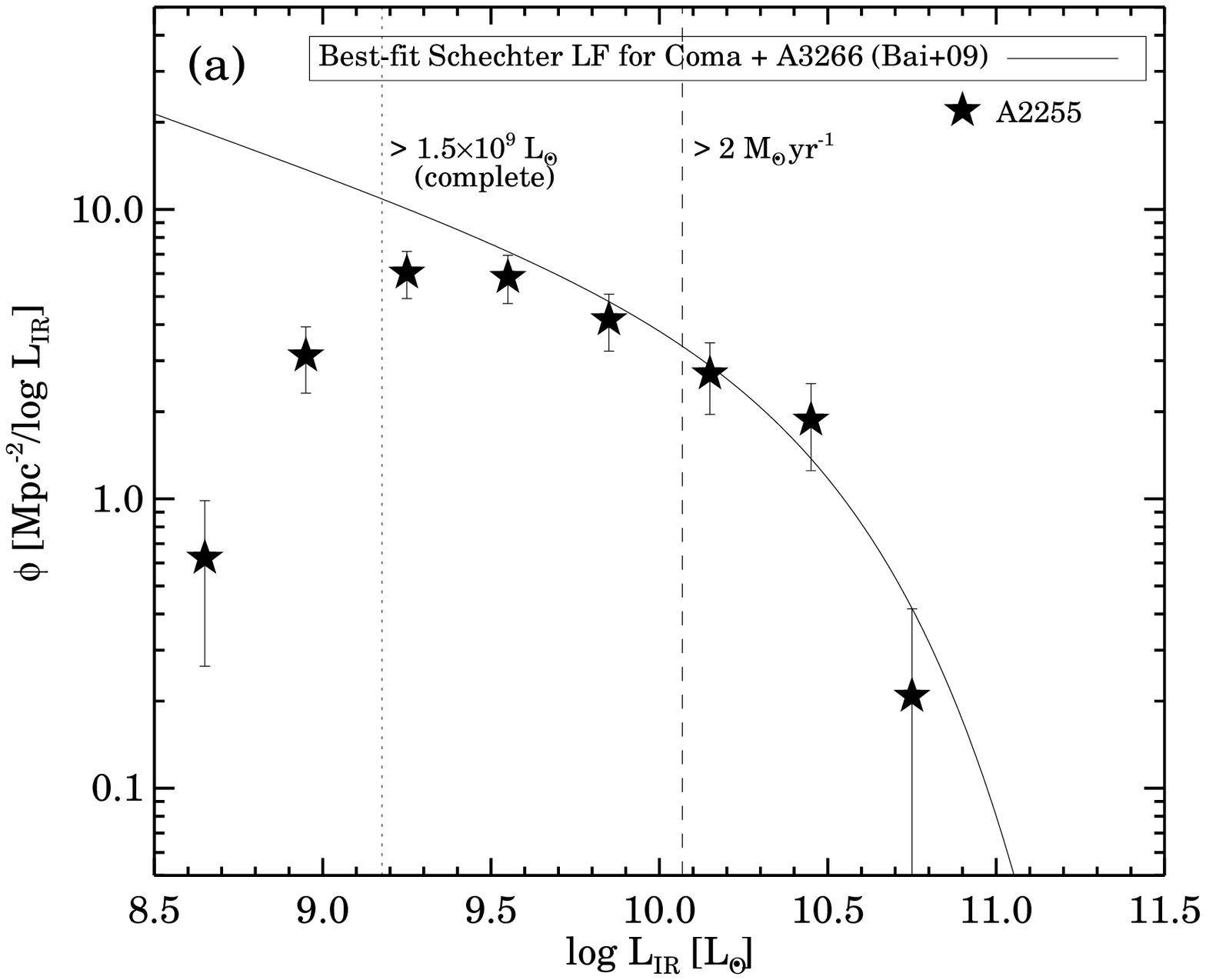}{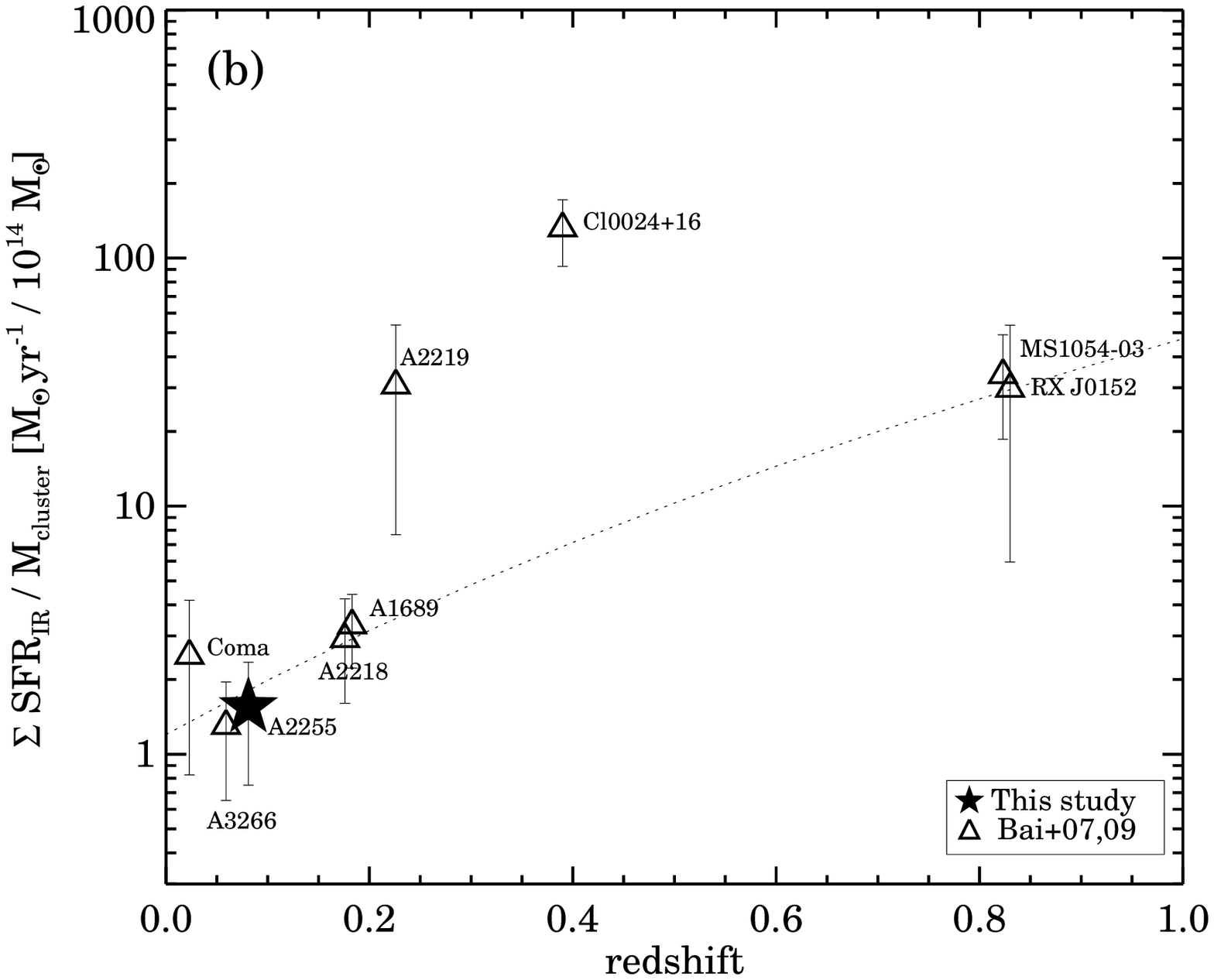}
  \caption{ \label{fig:irlf}
  \textit{Left}:
  IR luminosity function of A2255. The solid line is the best-fit
  Schechter luminosity function derived using composite of Coma cluster
  and A3266 (Bai et al. 2009), with $\alpha=-1.4$ (fixed) and
  $\mbox{log} L_{IR}^* = 10.49~L_{\odot}$. The error bars indicate
  Poisson error in each bin. The filled stars are points from
  galaxies with ``strong'' MIR-excess.
  The dotted vertical line represents $L_{IR} > 1.5\times10^9~L_{\odot}$
  -- as described in Figure \ref{fig:irlimit}, our data is considered 
  to be complete at this limit.
  The dashed vertical line represents limit of SFR$>2~M_{\odot} \mbox{yr}^{-1}$,
  used to calculate $\sum$SFR in Figure \ref{fig:irlf}b.
  \textit{Right}:
  Total SFR of A2255 normalized to the cluster mass (i.e., specific 
  SFR). Our result is marked as filled star, and is compared with 
  results for other galaxy clusters from previous studies. 
  Open triangles are data points drawn from Bai et al.(2007, 2009).
  The dotted line represents $(1+z)^{5.3}$, an evolutionary trend of
  specific SFR for galaxy clusters (Bai et al. 2009).
  }
  \epsscale{1.0}
\end{figure*}

  We present the derived IR luminosity function
 with the Poisson error bars in Figure \ref{fig:irlf}a/ 
  The overplotted solid line is the best-fit Schechter luminosity function
 for Coma and A3266 composite (Bai et al. 2009), with parameters $\alpha=-1.41$
 and log\,$L_{IR}^* ~(L_{\odot}) = 10.49$. 
  The IR luminosity function of A2255 is consistent
 with those of Coma and A3266
 in terms of both the shape and the normalization
 above the completeness limit of IR luminosity
 ($L_{IR}\sim~1.5\times10^9~L_{\odot}$, Figure \ref{fig:irlimit}).

  Comparing the IR luminosity functions of Coma and A3266, Bai et al.(2009)
 suggested a possibility that local galaxy clusters share a universal
 IR luminosity function. 
 A2255 is close to the Coma cluster in terms of cluster mass 
 ($\sim10^{15}~M_{\odot}$, Burns et al. 1995; Lokas \& Mamon 2003),
 but has a slightly lower mass than A3266
 ($\sim3.3\times10^{15}~M_{\odot}$, Bai et al. 2009).
  In terms of the dynamical status, A2255, A3266, and the Coma 
 cluster all share similar properties -- these galaxy clusters are
 suggested to be in the late phase of cluster-cluster merger 
 (e.g., Watanabe et al. 1999; Sauvageot, Belsole, \& Pratt 2005).
  The consistency of A2255 IR luminosity function with those of Coma and 
 A3266 supports the idea of universal IR luminosity function in local galaxy
 clusters (e.g., Bai et al. 2009), at least when they are in the late stage 
 of cluster-scale merger. Yet it should be noted that the number of galaxy
 clusters studied is small, and most galaxy clusters studied to date are 
 biased to dynamically unrelaxed system.

  With the derived IR luminosity function, we calculate the ``total'' SFR 
 in A2255 by integrating the IR luminosity function. In the luminosity range of
 $L_{IR}>1.5\times10^9~L_{\odot}$, i.e. $SFR(IR)>0.25~M_{\odot}$ yr$^{-1}$, 
 the integration of the IR luminosity function yields
 $\sum_{>0.25~M_{\odot} yr^{-1}}^{LF} SFR = 115~M_{\odot}$ yr$^{-1}$ for A2255. 
  This is consistent with the summation of SFRs of ``individual'' member
 galaxies, $\sum_{>0.25~M_{\odot} yr^{-1}} SFR = 131~M_{\odot}$ yr$^{-1}$.

  In the SFR summation of individual star-forming galaxies 
 ($\sum_{>0.25~M_{\odot} yr^{-1}} SFR = 131~M_{\odot}$ yr$^{-1}$), we split 
 the contribution to the SFR from spectroscopic member galaxies and 
 photometric member galaxies in order to see the effects from the cluster
 membership determination.
 Among $131~M_{\odot}$ yr$^{-1}$, the member galaxies with
 spectroscopic redshifts contribute $103~M_{\odot}$ yr$^{-1}$,  
 and the member galaxies with photometric redshifts
 contribute $28~M_{\odot}$ yr$^{-1}$.
  The contribution from galaxies with photometric redshifts
 (i.e. optically faint member galaxies) is about 20\% : therefore,
 even at the very unlikely case of all photometric
 members turn out to be non-members, it does not have a strong effect on the
 derived total SFR. 

  In order to compare the star formation activity of A2255
 with other galaxy clusters, we use a ``specific SFR''
 of galaxy cluster -- the total SFR normalized
 by the cluster virial mass ($\sum SFR / M_{cl}$).
 The result is marked as a filled star in Figure \ref{fig:irlf}b.
  The $\sum SFR/\mbox{M}_{cl}$ value is calculated by integrating
 SFRs of galaxies with SFRs larger than $2~M_{\odot}$ yr$^{-1}$
 and within $0.5 ~r_{200}$ radius, in order to facilitate 
 the comparison with the mass-normalized SFR of other clusters 
 (other points; Bai et al. 2007, 2009).
  The points for Coma, A3266, MS1054-03, and RX J0152 
 (open triangles in Figure \ref{fig:irlf}b) are from the \textit{Spitzer}
 MIPS 24\,$\mu$m studies (Bai et al. 2007, 2009), while the points
 for A2218, A1689, A2219, and Cl0024+16 are from ISO observation
 (Bai et al. 2007).
  The sum of A2255 SFR is
 $\sum SFR_{IR} (>2~M_{\odot} \mbox{yr}^{-1}, \mbox{within} ~0.5~r_{200}) = 15.5~M_{\odot} \mbox{yr}^{-1} $. 
  We use $r_{200} = 2.10$\,Mpc and $M_{cl} = 0.45-1.3\times10^{15}~M_{\odot}$ 
 from previous studies (Neumann 2005; Burns et al. 1995; Feretti et al. 1997).
  The derived specific SFR of A2255 is comparable to those of four other
 clusters at $z<0.2$. On the other hand, compared to
 two intermediate-redshift clusters, A2219 and Cl0024+16, the specific 
 SFR of A2255 is more than an order of magnitude lower. 
 From the study of the total SFR in A2255, we find no evidence
 that the overall star formation activity is enhanced in A2255 compared to
 other clusters at low to intermediate redshifts.

\section{ Environmental Dependence of Star Formation and Galaxy Transformation } 

  We investigate the correlation between the galaxy evolution and
 their environment to see the role of the environment in the evolution
 of A2255 member galaxies. 
  Firstly, we define the parameters that represent local environment
 of galaxies. Secondly, we examine how different levels of MIR-excess 
 correlates with the environment.

 \subsection{Substructures and Kinematic Characteristics in A2255}

\begin{figure*}
 \epsscale{0.8} \plotone{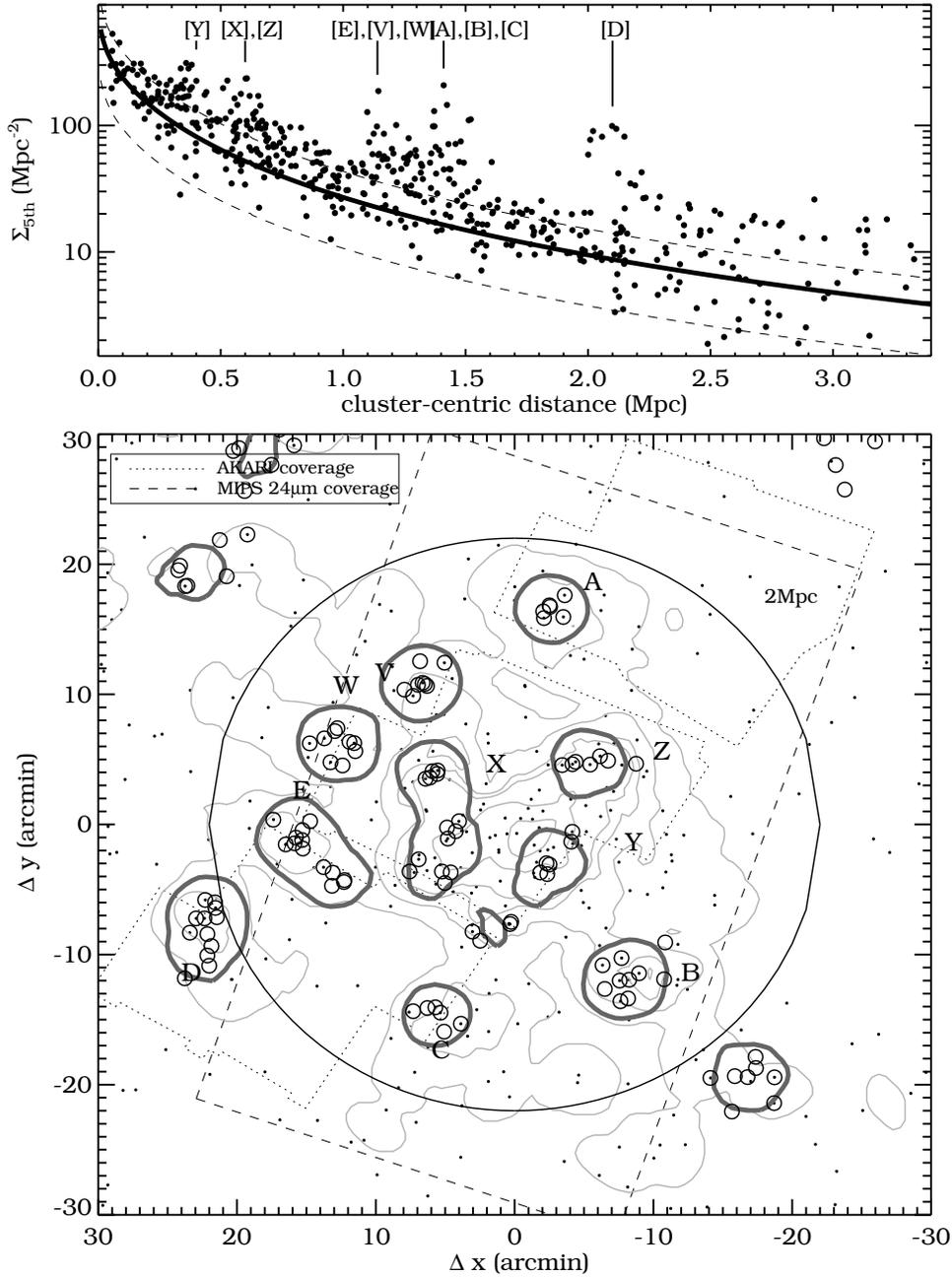}
 \caption{ \label{fig:substructure}
  \textit{Top} : Local galaxy surface density as a function of clustercentric
 distance. Local galaxy surface density is expressed as
 $\Sigma_{5th}$, galaxy number density within a circle
 with a radius of a distance to the fifth-nearest galaxy. The overplotted
 thick solid line has a form of
 two-dimensional projection of NFW profile for dark matter halos
 (Navarro, Frenk, \& White 1997; projection form adopted from
 El\'iasd\'ottir \& M\"oller 2007). The best-fit line is derived
 by excluding the outliers iteratively. The dashed lines indicate
 $ \pm \sigma_{rms}$ from the best-fit solid line.
 There are several peaks of galaxies with $\Delta|\Sigma_{5th}| > \sigma_{rms}$,
 i.e. galaxies showing larger local surface density compared to
 the expected surface density from a relaxed system.
 These peaks of galaxies are defined as ``substructure''s A-E, and V-Z
 in the bottom panel. 
  \textit{Bottom} : Spatial distribution of A2255 member galaxies. 
 The overlaying contours are number density contours of galaxies
 with spectroscopic redshifts of $0.068<z<0.095$. Open circles indicate
 ``galaxies at high-density region'', which lies in the peaks of the 
 top panel. 
 }
\end{figure*}

  We use the local surface density of galaxies and clustercentric radius
 as indicators of environment. 
  The local surface density of galaxies is expressed as $\Sigma_{5th}$,
 the density of member galaxies (with either spectroscopic or photometric
 redshifts) within a circle whose radius is a distance to the 5th-nearest
 galaxy in comoving scale. Galaxies brighter than $r\sim19$ mag are used
 for calculating the local density. If a galaxy cluster is a simply
 relaxed system, the local surface density would monotonically decrease
 as the clustercentric distance increases.
  However for clusters like A2255, this is not the case. 
  We identify substructures with densities higher than the local average
 densities to be areas that deviates from a smooth projected number density
 represented by the NFW function (Navarro, Frenk, \& White 1997).
  To fit the relation
 between the clustercentric distance vs. the observed $\Sigma_{5th}$,
 we use a two-dimensional projected form of the NFW function 
 (El\'iasd\'ottir \& M\"oller 2007;
 see the dashed line in Figure \ref{fig:substructure}a) : 

 \begin{equation}
  \begin{array}{l}
  \Sigma_{nfw} = \frac{2 r_s \delta_c \rho_c}{1 - X^2} \left(1 - \frac{2}{\sqrt{1-X^2}} \mbox{arctanh} \sqrt{\frac{1-X}{1+X}}   \right) ,~~~~~~ X<1  \\
  \Sigma_{nfw} = \frac{2 r_s \delta_c \rho_c}{X^2 - 1} \left(1 - \frac{2}{\sqrt{X^2-1}} \mbox{arctan} \sqrt{\frac{X-1}{1+X}}   \right) ,~~~~~~ X>1
  \end{array}
 \end{equation}

  In this equation, $X$ indicates $R_{cl}/r_s$, where $R_{cl}$
 is the clustercentric radius. The parameters $r_s, \delta_c$ and $\rho_c$
 represent the scale radius, the characteristic overdensity,
 and the critical density. Our best-fitted parameters are 
 $r_s = 0.68$\,Mpc and $\delta_c \rho_c =127$\,Mpc$^{-2}$. 
 The galaxies above $\sigma_{rms}$ from the fitted line is considered
 as galaxies at substructures with locally high density of galaxies.
  The coordinates of these galaxies are plotted as open circles
 in Figure \ref{fig:substructure}b, over the distribution of
 all member galaxies in the Y03 catalog. 
   The overplotted thick lines are contours of ``galaxies at high density
 environment'', and we define these peaks as substructures. This means
 it is not necessary that all substructure galaxies have high $\Sigma_{5th}$,
 although most of them do. The locations of these structures are consistent
 with galaxy number density contour.
  The coordinates of the substructures marked
 in Figure \ref{fig:substructure}b are --
 (17:13:44, 64:14:42) for substructure ``V'',
 (17:14:48, 64:10:00) for ``W'',
 (17:13:10, 64:05:00) for ``X'',
 (17:12:20, 64:00:00) for ``Y'', and
 (17:11:40, 64:08:00) for ``Z''.
  Substructures A, B, C, D, and E are consistent with the substructures
 defined in Yuan, Zhou, \& Jiang (2003) with the same alphabets. They
 first defined the substructures
 from the peak of number density contour ($>0.15$ galaxies arcmin$^{-2}$,
 see Figure 5 of Yuan, Zhou, \& Jiang 2003).
 By studying the velocity distribution of galaxies in the substructures,
 these authors confirm that substructures A, B, and C have
 different velocity components compared to the main cluster, while
 substructures D and E might be a result of a projection effect.
  Our MIR observation is limited within the central $\sim2$\,Mpc
 radius of A2255. Consequently, the substructures are all defined within this 
 $\sim2$\,Mpc radius circle (about virial radius of the cluster). 

\begin{figure*}
 \epsscale{0.9} \plotone{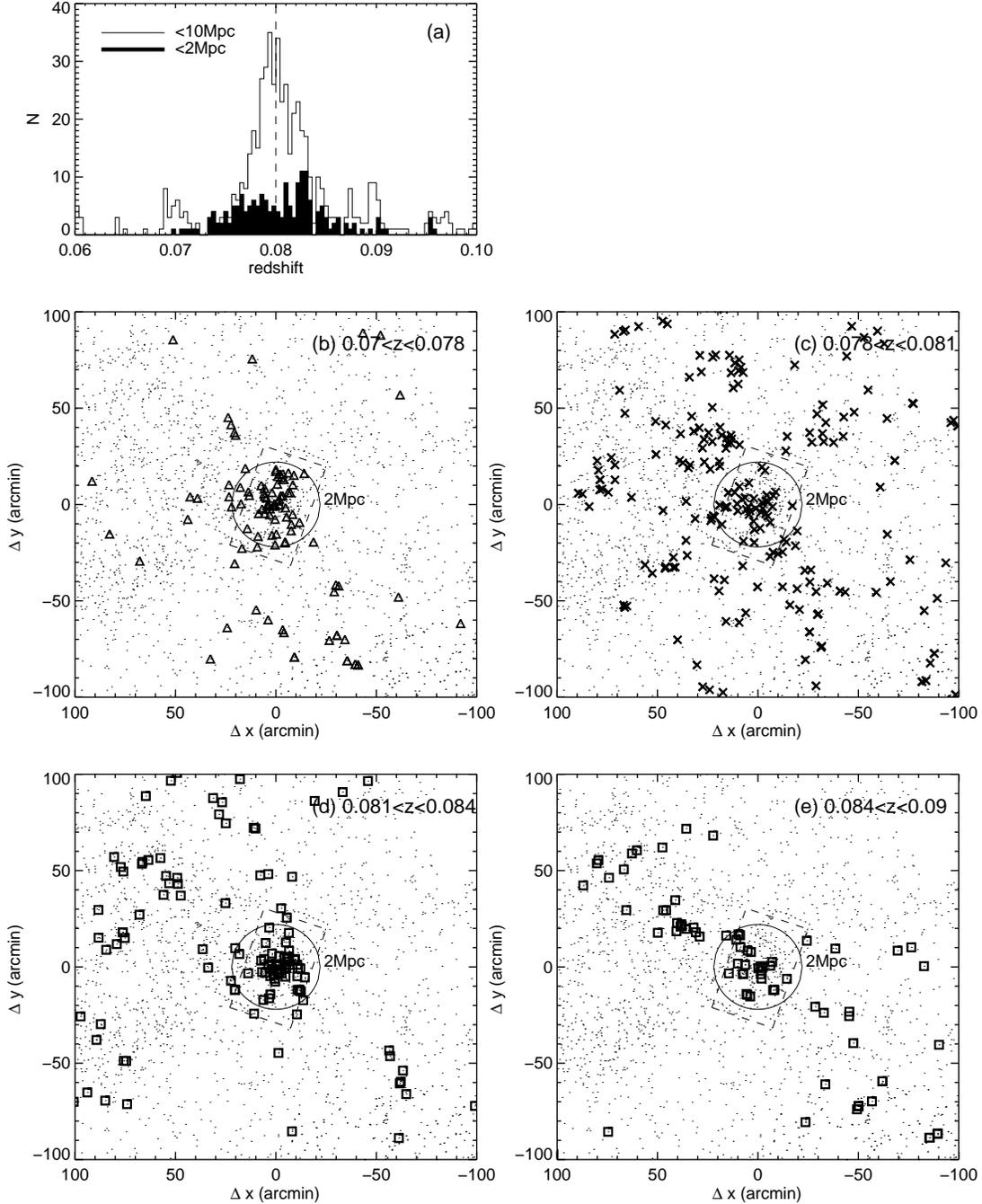}
  \caption{ \label{fig:largescale} (a) The spectroscopic redshift 
  distribution of A2255 member galaxies and galaxies in the vicinity
  of A2255. While galaxies
  within $<10$\,Mpc distance from the A2255 center show the redshift
  distribution peaked at $z\sim0.08$ (\textit{dashed} line), the
  redshift distribution of A2255 member galaxies is skewed to 
  have two distinct velocity peaks ($z_1=0.077$ and $z_2=0.082$).
  (b) Spatial distribution of galaxies at $0.07<z<0.078$.
  The circle indicates 2\,Mpc radius circle, and the \textit{dotted/dashed} 
  line shows the fields of view of the AKARI IRC/\textit{Spitzer}
  MIPS. 
  (c) Spatial distribution of galaxies at $0.078<z<0.081$.
  (d) Spatial distribution of galaxies at $0.081<z<0.084$.
  (e) Spatial distribution of galaxies at redshifts $0.084<z<0.09$.
  }
\end{figure*}

  Besides of the substructures, we also investigate the large 
 scale distribution of galaxies around A2255, since A2255 is known
 to be a member of the North Ecliptic Pole (NEP) supercluster
 (e.g., Mullis et al. 2001). Understanding the large scale  
 distribution of galaxies will be helpful to see how the distribution
 of galaxies in A2255 is connected to the features such as walls 
 and filaments.  
 Figure \ref{fig:largescale}a shows the redshift distribution 
 of galaxies located in $3\times3$\,deg$^2$ region surrounding A2255.
 The spectroscopic redshifts are obtained from SDSS DR7 
 (Abazajian et al. 2009). While the solid histogram shows the 
 redshift distribution of all the galaxies within a circle with
 10\,Mpc radius centered on A2255, the filled histogram indicates
 the redshift distribution of galaxies within 2\,Mpc distance 
 from the cluster center.  The large scale
 analysis shows strong redshift peak at $z\sim0.08$ which can be 
 attributed to the presence of A2255 and the NEP supercluster. 
 On the other hand, in the central region within the clustercentric
 distance of 2\,Mpc, the redshift distribution is far different from 
 a single Gaussian distribution : 
  the possibility that the distribution of galaxies located within the
 central 2\,Mpc follow a single Gaussian function is only 40\% 
 by Kolmogorov-Smirnov test.
 Instead, the redshift distribution is well described by a
 sum of two redshift components with $z_1 = 0.077\pm0.0027$ 
 and $z_2 = 0.082\pm0.0017$, with the lack of galaxies being 
 at $z\sim0.08$.
  Figure \ref{fig:largescale}b, c, d, and e show the
 spatial distribution of galaxies at different redshift bins. 
 The galaxies at the blue velocity component (b; $0.07<z<0.078$)
 and the red velocity component (d; $0.081<z<0.084$) are 
 concentrated in the cluster center. The redshift range of 
 $0.078<z<0.081$, where only few of A2255 cluster members 
 reside in, consists of galaxies that are distributed like 
 a flat sheet at $z\sim0.08$ (Figure \ref{fig:largescale}c). 
 At the reddest velocity tail (e; $0.084<z<0.09$), galaxies are
 distributed along a filamentary structure that pass through A2255. 
  The two velocity groups ($0.07<z<0.078$ and $0.081<z<0.084$) 
 are virtually superimposed in the sky.
  The velocity dispersion of two groups are 800\,km s$^{-1}$ (blue)
 and 500\,km s$^{-1}$ (red). Using simple virial theorem, the 
 masses of each component are $1.7\times10^{15}~M_{\odot}$ (blue)
 and $6.7\times10^{14}~M_{\odot}$ (red) when virial radius is 2\,Mpc
 for each group. According to the criteria for gravitationally bounded 
 two-body problem (Beers, Geller, \& Huchra 1982; Tran et al. 2005), 
 the probability that these two components are gravitationally
 bounded is when the following equation is satisfied. 

 \begin{equation}
    V_r^2 R_p \le 2 G M ~ sin^2\alpha ~ cos\alpha
 \end{equation}

  In this equation, $V_r$ is the relative velocity between the
 two groups (1500\,km s$^{-1}$), $R_p$ is the projected separation
 (less than $\sim100$\,kpc since two components are nearly superimposed), 
 $M$ is the total mass of the system and $\alpha$ is the projected 
 angle with respect to the plane of the sky. The equation is 
 valid  for the projection angle range of $6\arcdeg<\alpha<89\arcdeg$,
 thus the probability that the two groups are bounded is more than 90\%.
 Therefore, A2255 is clearly thought to be ``merging'' galaxy cluster
 with two different velocity peaks. 

\subsection{Environmental Dependence of Star Formation}

\begin{figure*}
 \epsscale{1.0} \plotone{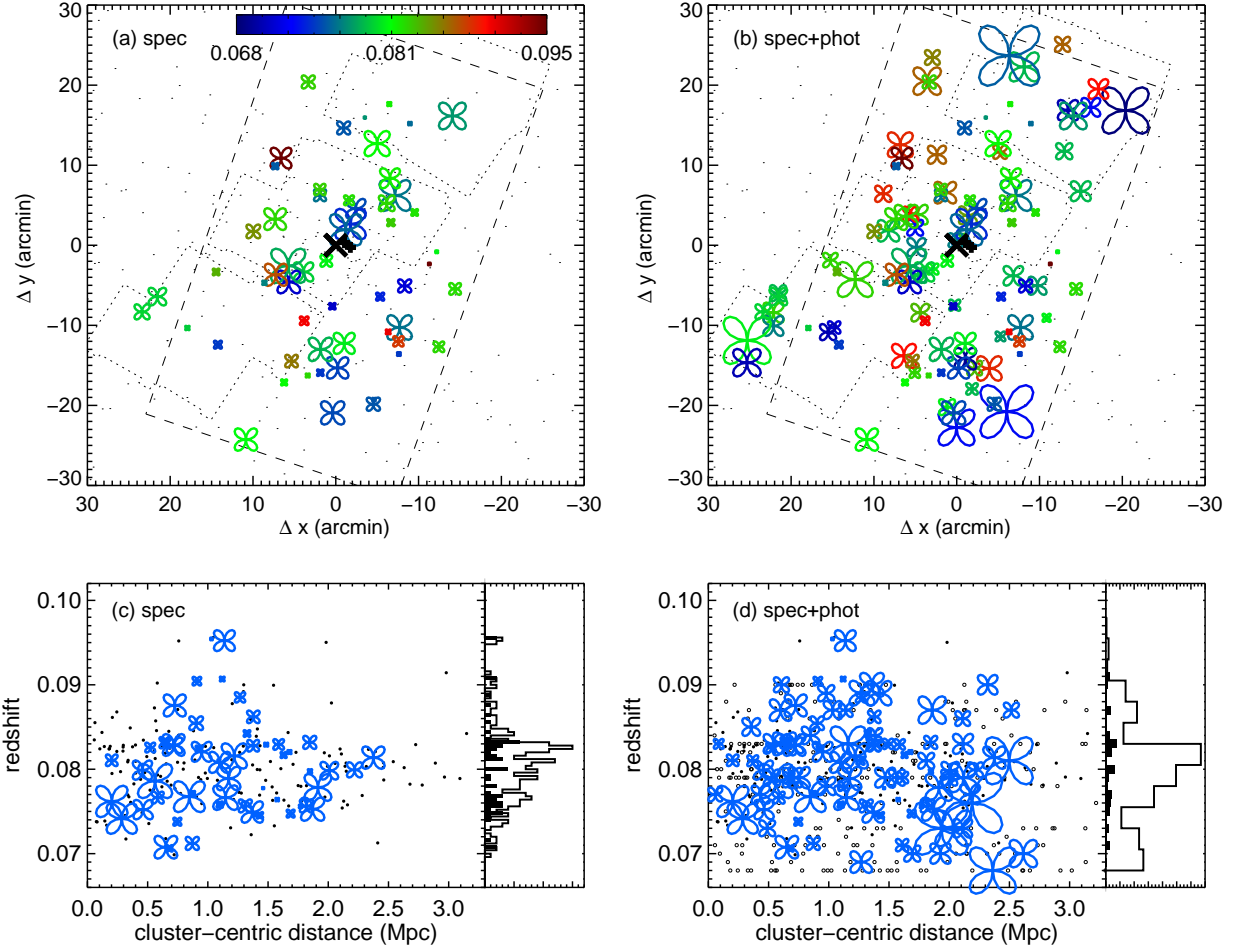}
  \caption{ \label{fig:spatial_sf}
 (a) Two-dimensional spatial distribution of star-forming galaxies
 in A2255. Galaxies with spectroscopic redshifts are plotted as 
 clovers. The symbol color denotes the redshift of a galaxy as
 indicated in the color bar, and the symbol size is proportional
 to the specific SFR.
 (b) Same as in (a), galaxies with either spectroscopic or photometric
 redshifts are plotted.
 (c) The clustercentric distance and velocity distribution of 
 star-forming galaxies in A2255. The points plotted as small dots
 are all member galaxies with spectroscopic redshifts defined in Y03 catalog.
 Overplotted clovers are star-forming galaxies, with symbol size 
 proportional to the specific SFR. On the right, we present the
 histogram showing the velocity distribution of galaxies -- 
 solid histogram for all member galaxies, and filled histogram
 for star-forming galaxies only.  
 (d) Same as in (c), galaxies with either spectroscopic or photometric
 redshifts are plotted. 
 }
 \epsscale{1.0}
\end{figure*}

  Figure \ref{fig:spatial_sf}a and \ref{fig:spatial_sf}b show the projected
 two-dimensional spatial distribution of star-forming galaxies in A2255.
 The color of each point indicates the redshift, and the size of each
 point is proportional to the specific SFR (SSFR). 
  The figures show that galaxies with high SSFR are located preferentially
 in the outer region of the cluster (Figure \ref{fig:spatial_sf}a, b),
 reflecting the well-known morphology-density relation. There are a few
 galaxies with high SSFR near the cluster center, but they have mostly
 redshifts much lower than the mean redshift of cluster (blue points 
 in the center of Figure \ref{fig:spatial_sf}a). 
  The $R_{cl}-v$ plot of star-forming galaxies (Figure \ref{fig:spatial_sf}c, d)
 clearly confirms the trend, that galaxies with high SSFR are either located
 at the outer region of A2255 or at the blue velocity peak. 
 
\begin{figure}
 \epsscale{0.8} \plotone{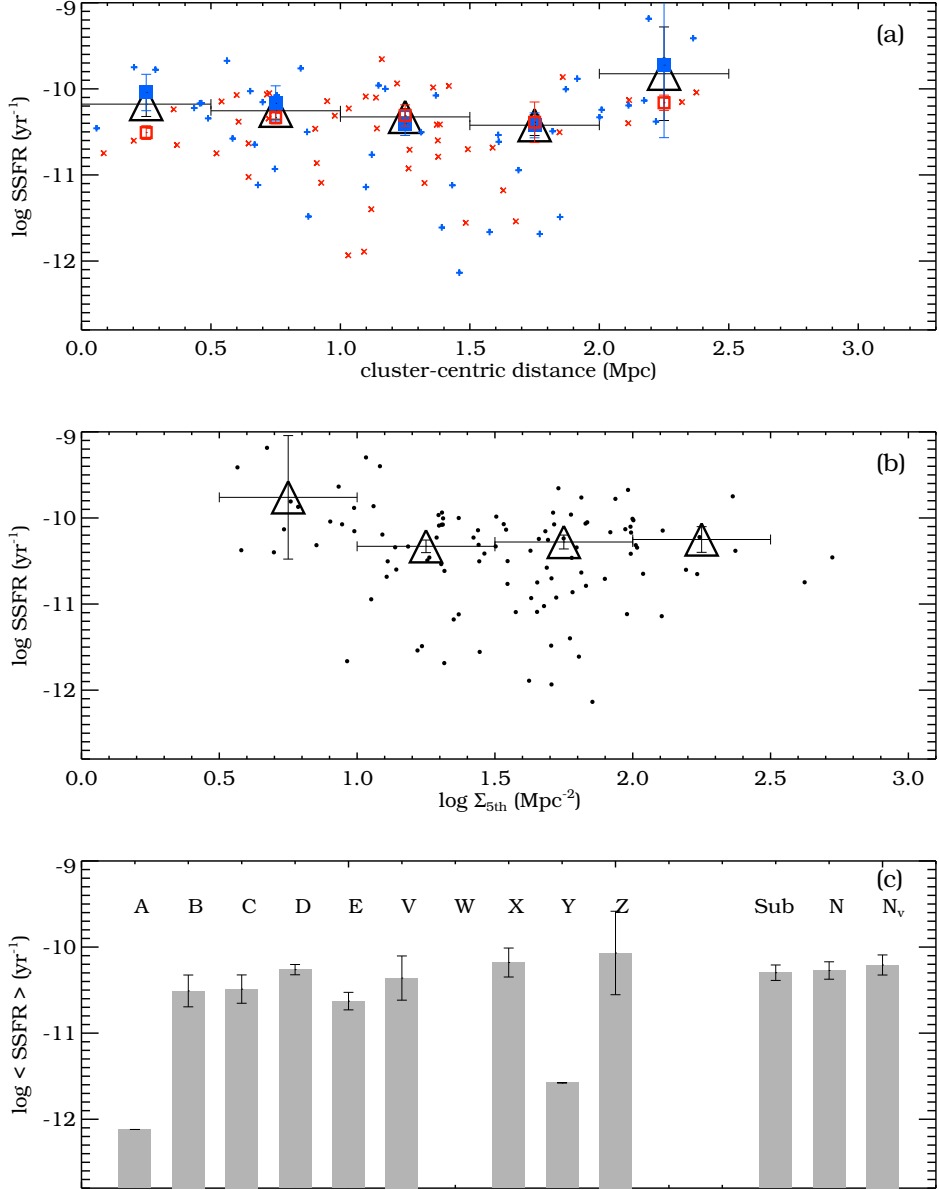}
  \caption{ \label{fig:env_ssfr}
  \textit{Top} : Specific SFR (SSFR) of galaxies as a function of 
  clustercentric radius (\textit{dots}). The \textit{triangles} 
  indicate the average SSFR at each bins of clustercentric distance,
  with error bars produced by bootstrapping. 
  The \textit{Blue} \textit{filled} squares represent average SSFR from 
  star-forming galaxies at blue velocity peak ($z<0.08$), 
  and the \textit{red} \textit{open} squares represent average SSFR from
  star-forming galaxies at red velocity peak ($z>0.08$).
  Each blue and red cross indicates individual galaxy with either 
  blue or red velocity.  
      \textit{Middle} : SSFR of galaxies as a function of local surface
  density of galaxies ($\Sigma_{5th}$). 
      \textit{Bottom} : The average SSFR for galaxies brighter than 
  $r=19.0$ mag in each substructure (A-Z), in all substructure (Sub),
  and outside the substructures (N). 
  The error bars are gain derived through bootstrapping. Points for
  substructure A and Y are generated by only one galaxy, therefore
  we do not mark error bars. 
   For galaxies outside the substructures,
  average SSFR of those with velocities at the blue/red end of the velocity
  distribution is marked as N$_v$ ($z<0.075$ or $z>0.085$).
  }
\end{figure}

  In order to analyze the distribution of galaxies with high SSFR in more 
 detail, we investigate SSFR as a function of environmental parameters in 
 Figure \ref{fig:env_ssfr}. Figure \ref{fig:env_ssfr}a shows $R_{cl}$ vs. SSFR
 plot. The average SSFR does not change much as a function of $R_{cl}$, i.e. 
 there is no clear sign of enhancement in SSFR as a function of $R_{cl}$.
  Figure \ref{fig:env_ssfr}b shows how SSFR changes as a function of 
 local galaxy surface density, $\Sigma_{5th}$. 
  Previous works of intermediate redshift clusters suggested that SSFR is
 enhanced at the intermediate density region of clusters and the enhanced
 activity is related to infall of galaxies into the gravitational potential
 of clusters (e.g., Koyama et al. 2010). For A2255, we find that there is 
 no clear enhancement of SSFR from surface density of $\Sigma_{5th}=0.5-2.5$\,Mpc$^{-2}$.
  There is a slight increase in lower density region, at $\Sigma_{5th}=0.5$\,Mpc$^{-2}$,
 yet the amount of difference is less than a factor of 1.5. Therefore, we
 conclude that star formation is not enhanced in dense regions of A2255. 

  We also examine if star formation activities are enhanced at substructures
 as suggested from studies of other clusters (e.g., Koyama et al. 2010).
 Figure \ref{fig:env_ssfr}c shows the average SSFR for the substructures
 identified in Section 5.1, in comparison to galaxies in the other areas
 of the cluster. Errors for the average SSFR are the standard deviations
 of SSFR distribution for each case. 
  We find that there is little difference between the average SSFR of
 galaxies in substructures (Sub) and not in substructures (N). 
 Several substructures (A, W, and Y) contain very few star-forming galaxies 
 and the average SSFRs are low. From this, we conclude that SSFR is not 
 particularly enhanced in substructures even if some galaxies in 
 substructures show high SSFR compared to global average.

  Although we find no convincing evidence for the enhancement of SSFR
 at a particular environment (except that SSFRs are higher at the outer region),
 we find an interesting trend in the distribution of star-forming galaxies 
 in the inner region. As seen in Figure \ref{fig:spatial_sf}c, 
 the redshift distribution of star-forming galaxies (filled histogram)
 is bimodal : a blue velocity peak at $\langle z \rangle=0.077$ and
 a red velocity peak at $\langle z\rangle=0.082$ (Note that this analysis
 is only possible for samples with spectroscopic redshifts since the
 accuracy of photometric redshifts is too low). At $R_{cl}<0.5$\,Mpc,
 the majority of galaxies with high SSFR belong to the blue velocity peak. 
  We show the average SSFR of galaxies at the blue velocity peak and at the red
 velocity peak in Figure \ref{fig:env_ssfr}a. Figure \ref{fig:env_ssfr}a
 indicates that the average SSFR of galaxies at the blue velocity peak ($z<0.08$) is 
 higher than that of galaxies at the red velocity peak ($z>0.08$) by a factor of
 $\sim3$, at the innermost region with $R_{cl}<0.5$\,Mpc.
  By randomly selecting and assigning galaxies at $R_{cl}<0.5$\,Mpc 
 to either blue or red peaks, we find that the chance probability of
 finding this observed difference in the average SSFR is less than 5\%. 
 This trend appears to continue out to $R_{cl}\sim1.0$\,Mpc, but the trend
 is not statistically significant compared to what we found at $R_{cl}<0.5$\,Mpc.

\subsection{Environmental Dependence of Galaxy Transformation}

\begin{figure*}
 \epsscale{1.0} \plotone{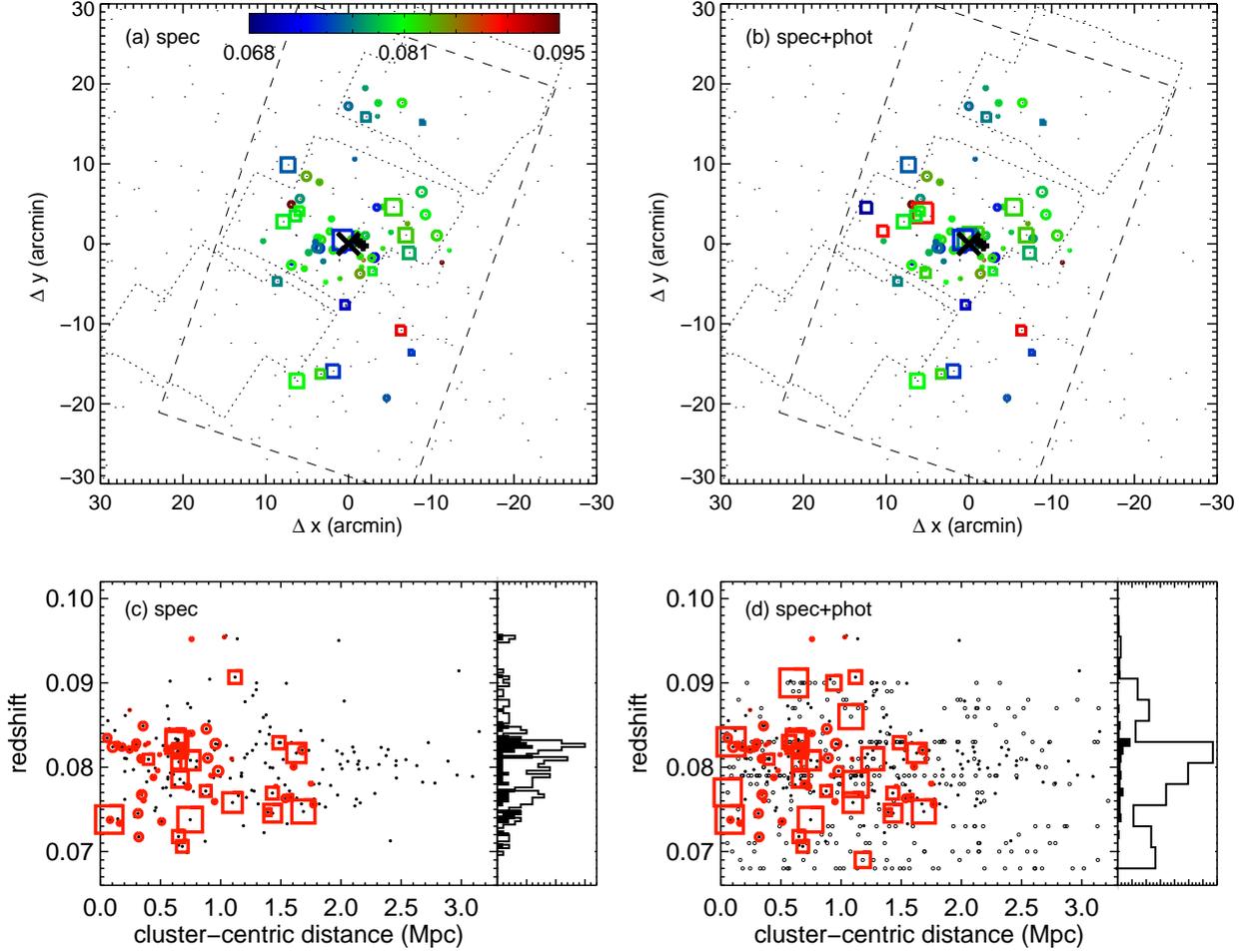}
  \caption{ \label{fig:spatial_ell}
 (a) Two-dimensional spatial distribution of weak/intermediate
 MIR-excess galaxies in A2255. The \textit{circles} indicate 
 weak MIR-excess galaxies, and the \textit{squares} indicate
 intermediate MIR-excess galaxies. The symbol color denotes
 the redshift of a galaxy as indicated in the color bar,
 and the symbol size is proportional to the $N3-S11$ color.
 In case of mag($z$)$-$mag(24\,$\mu$m) color, the value is converted
 to $N3-S11$ according to the relation in Figure \ref{fig:mircolor}b.
 (b) Same as in (a), galaxies with either spectroscopic or photometric
 redshifts are plotted.
 (c) The clustercentric distance and velocity distribution of 
 weak/intermediate galaxies in A2255. The points plotted as small dots
 are all member galaxies with spectroscopic redshifts defined in Y03 catalog.
 Overplotted are weak(\textit{circle})/intermediate(\textit{square})
 MIR-excess galaxies, with symbol size proportional to the $N3-S11$ color.
 On the right, we present the histogram showing the velocity distribution
 of galaxies -- solid histogram for all member galaxies, and filled histogram
 for weak/intermediate MIR-excess galaxies only. 
 (d) Same as in (c), galaxies with either spectroscopic or photometric
 redshifts are plotted.
 }
 \epsscale{1.0}
\end{figure*}

\begin{figure}
 \epsscale{0.8} \plotone{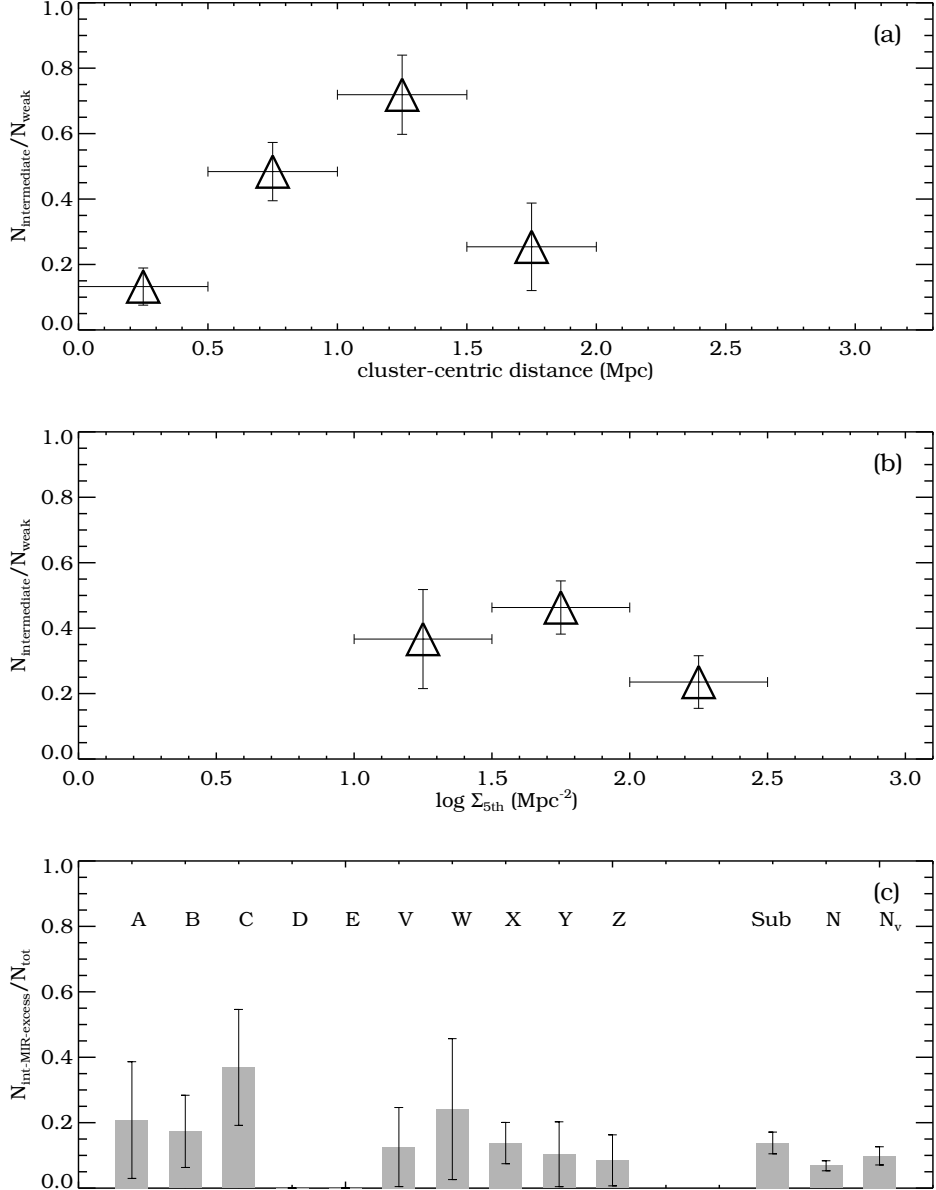}
  \caption{ \label{fig:env_mirx}
       \textit{Top} : The ratio between the number of intermediate MIR-excess
  galaxies and the number of weak MIR-excess galaxies, as a function of
  clustercentric radius. The values are derived using $r<19.0$ mag galaxies.
  The error bars are derived through bootstrapping.
       \textit{Middle} : The ratio between the number of intermediate MIR-excess
  galaxies and the number of weak MIR-excess galaxies, as a function of
  local surface density of galaxies. Again, the values are derived using $r<19.0$ mag
  galaxies, and the error bars indicate bootstrapping errors. 
       \textit{Bottom} : The fraction of intermediate MIR-excess galaxies
  among the `total' galaxies in substructures, and outside the substructures
  ($N_{tot}$ indicates total number of galaxies brighter than $r=19.0$ mag 
  in each substructures, or outside the substructures, including all levels of
  MIR-excess galaxies and MIR non-excess galaxies).
  The marks for substructures A-Z and galaxies outside the substructures
  (N), galaxies outside the substructures with extreme velocities (N$_v$)
  are the same as in the Figure \ref{fig:env_ssfr}.
  }
\end{figure}

\begin{figure}
 \epsscale{1.0} \plotone{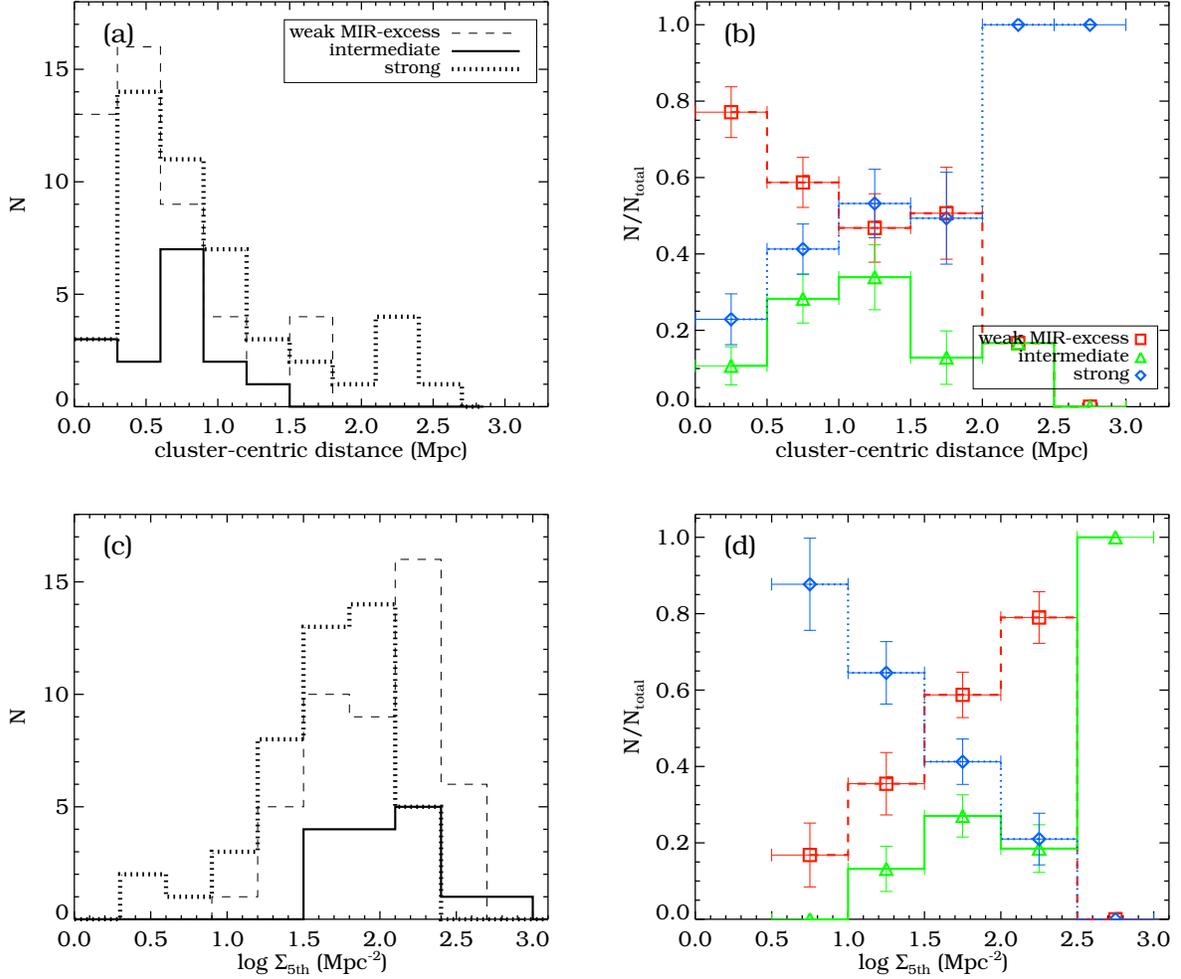}
 \caption{ \label{fig:env}
  (a) The distribution of MIR-excess galaxies of different levels 
  as a function of clustercentric distance. The galaxies used in this
  plot are selected above the completeness limit of member identification
  (i.e., $r<19.0$ mag). \textit{Dashed} line indicates weak MIR-excess
  galaxies, \textit{solid} line indicates intermediate MIR-excess
  galaxies, and \textit{dotted} line indicates strong MIR-excess galaxies. 
   The same legend applies to plot (b)-(d). 
  (b) The fraction of weak/intermediate/strong MIR-excess galaxies  
  among the total galaxies as a function of clustercentric distance. 
  The error bars are derived through bootstrapping. 
  (c) The distribution of MIR-excess galaxies of different levels 
  according to the local surface density of galaxies ($\Sigma_{5th}$). 
  (d) The fraction of MIR-excess galaxies according to the local surface
  density of galaxies. The error bars and shifts of the points are the 
  same as in panel (b). 
  }
\end{figure}

  As we discussed in Section 3.4, intermediate MIR-excess galaxies are 
 probes of galaxy transformation from star-forming galaxies (strong MIR-excess)
 to quiescent galaxies (weak MIR-excess).
  By studying the environment of intermediate MIR-excess
 galaxies, we can gain insight on where in the cluster the galaxy transformation
 is taking place and what physical mechanisms govern the transformation. 
  The two-dimensional spatial
 distribution of weak/intermediate MIR-excess galaxies is plotted in 
 Figures \ref{fig:spatial_ell}a and b. The plots show that 
 weak MIR-excess galaxies populate near the cluster center,
 while intermediate MIR-excess galaxies prefer $R_{cl}>0.5$\,Mpc. 
  We show the trend in a more quantitative way in Figure \ref{fig:env_mirx},
 by presenting the number ratios of galaxies with different levels of
 MIR-excess as a function of the environmental parameters. 
  Figure \ref{fig:env_mirx}a illustrates the ratio of intermediate MIR-excess
 galaxies to weak MIR-excess galaxies as a function of the clustercentric radius.
 Figure \ref{fig:env_mirx}b illustrates the same number ratio as a function 
 of the local galaxy surface density. 
  These are clearer form of Figure \ref{fig:env}, showing the relative fraction
 of strong/intermediate/weak MIR-excess galaxies as a function of environment, 
 with emphasis on the weak and intermediate MIR-excess galaxies.
   Figure \ref{fig:env}b and d clearly shows that the weak MIR-excess galaxies
 dominate at the highest densities, and the strong MIR-excess galaxies 
 dominate at the lowest densities. The intermediate MIR-excess galaxies 
 start to show up at around 
 $\Sigma_{5th}\sim2.5$\,Mpc$^{-2}$ ($R_{cl}\sim0.5$\,Mpc), but their number 
 disappears quickly at $\Sigma_{5th}<1.5$\,Mpc$^{-2}$ ($R_{cl}\sim1.6$\,Mpc). 
  The relative fraction of intermediate MIR-excess to weak MIR-excess galaxies
 is the highest at the similar density/$R_{cl}$, $\Sigma_{5th}\sim1.5$\,Mpc$^{-2}$
 and $R_{cl}\sim1.3$\,Mpc (Figure \ref{fig:env_mirx}a, b). 
  From this, we conclude that galaxies in transformation (as represented by 
 intermediate MIR-excess galaxies) populate the intermediate density regions 
 (or the clustercentric distance of $R_{cl}=0.6-1.6$\,Mpc).
  This clustercentric radius corresponds to $0.3-0.8~R_{vir}$, 
 which exactly coincides with the location in clusters where the 
 ram pressure stripping is known to be efficient in quenching the
 star formation in galaxies by stripping their gas reservoir 
 based on the sphere of influence argument (Treu et al. 2003). 

  Like in case of star-forming galaxies, a significant number of 
 intermediate MIR-excess galaxies have redshifts at blue velocity 
 component (Figure \ref{fig:env_mirx}a). The difference with in the case
 of star-forming galaxies is that intermediate MIR-excess galaxies 
 at blue velocity peak are located at $R_{cl}>1$\,Mpc. 
  We also examine the correlation between substructures and the fraction
 of intermediate MIR-excess galaxies (Figure \ref{fig:env_mirx}c). 
 Like in the case of star-forming galaxies, there is no clear evidence 
 that intermediate MIR-excess galaxies are preferentially located in 
 the substructures.  Note that however, as we mentioned in 3.2,
 the observation of MIR-excess galaxies is limited by the depth of MIR images 
 -- especially at weak/intermediate MIR-excess for regions without $S11$-band 
 coverage. The substructures B, C, D, and E are such substructures,
 thus the reason we don't see weak/intermediate MIR-excess galaxies
 in these substructures may due to the shallow depth of the 24\,$\mu$m image. 
  Most of the galaxies in substructures B, C, D, and E have $r$-band magnitude
 of $16.5<r<18$ mag, therefore only objects with $z-24~\mu$m $>-1$, i.e.
 close to strong MIR-excess galaxies can be detected as intermediate MIR-excess
 galaxies.

\section{Discussion}

  The MIR data presented so far invoke several interesting points
 regarding the evolution of galaxies in a cluster environment like
 A2255, i.e. how cluster-scale dynamics correlates with the evolution of
 individual member galaxies. In Section 4.2, we show that the total SFR of
 A2255 is comparable with that of other galaxy clusters of similar mass and 
 dynamical stage, which implies that the cluster-scale merging activity 
 in A2255 does not enhance the star formation activity
 of member galaxies to the level of galaxies in some of the active 
 clusters at intermediate redshifts. 
  There is no clear evidence that substructures are closely 
 associated with galaxies with high SSFR 
 (Section 5.2, Figure \ref{fig:env_ssfr}c). 
 Star-forming galaxies are located at the outer part of the cluster
 ($R_{cl}>0.5$\,Mpc) or blue/red end of the velocity distribution 
 (Figure \ref{fig:spatial_sf}a, b). 
  While the dynamical status of A2255 is considered to be ``post-merger'' 
 from the previous X-ray observations
 -- no distinct multiple temperature peak, the temperature gradient
 elongated in east-west direction (Davis, Miller, \& Mushotzky 2003) --,
 our results are consistent with no enhancement of star formation 
 related to the cluster merger.

  This result is at odds with conclusions put forward in previous
 studies. Miller \& Owen (2003) suggested 
 that the fraction of radio selected star-forming galaxies
 are twice larger in A2255 compared to other galaxy clusters
 including Coma cluster. They referred this ``enhanced'' star formation
 in A2255 to the cluster-cluster merger, arguing that the alignment
 of star-forming galaxies in the north-south direction is a product of
 cluster-cluster merging in east-west direction. 
  However, as discussed earlier, we do not find such an alignment of
 star-forming galaxies nor more active star formation than the 
 Coma cluster. The discrepancy between their and our results can
 be understood as the following.
  Our MIR observation
 discover more star-forming galaxies in A2255 than Miller \& Owen (2003), 
 since we are probing the SFR much deeper than their SFR detection
 limit of $\sim2.5~M_{\odot}$ yr$^{-1}$. All the radio-selected 
 star-forming galaxies in Miller \& Owen (2003) are detected 
 in MIR if they are within either the AKARI or the \textit{Spitzer} fields.
  On the other hand, a number of strong MIR-excess galaxies 
 having comparable SFR with radio-selected star-forming galaxies are
 not detected in radio.
  Therefore we conclude that the difference between the MIR study
 and the radio study arises from the way how star-forming galaxies 
 are selected, and
 that the effect of the environment on star formation activities 
 requires a deep, multi-wavelength data not to miss galaxies
 with moderate SFR. In another work, Yuan et al.(2005) 
 suggested that cluster-cluster merging in A2255 had different 
 effect on star-forming galaxies with different morphologies, i.e.
 SFRs of late-type galaxies increase towards the center and 
 the SFRs of the early-type galaxies decrease towards the center
 using the SFR measured in optical spectroscopy, 
 concluding that the cluster-cluster merging activity triggered
 star formation in the central region of the cluster.
  However, our study shows no marked increase in the SSFR of galaxies 
 near the cluster center. Furthermore, star-forming galaxies near
 the cluster center are found to have significantly blueshifted 
 radial velocities with respect to the cluster mean redshift
 whose origin is discussed in more detail below. 

  We can gain another valuable insight on the galaxy evolution
 in a merging cluster from the velocity distribution of MIR-excess 
 galaxies. In Sections 5.2 and 5.3, we noted on a high fraction of
 star-forming galaxies and intermediate MIR-excess galaxies in the
 blue velocity peak and the lack of star-forming galaxies with 
 high SSFR at the redshift peak especially at $R_{cl}<0.5$\,Mpc.
  We discuss here how this interesting tendency can arise. 
  A simulation performed by Fujita et al.(1999) of a merging cluster
 illustrates how galaxies in the merging cluster evolve as the
 merging of the clusters proceed. According to their work, the 
 merging activity increases the ram-pressure, causing a more 
 efficient stripping of gas from galaxies in the cluster, hence
 causing a quenching of star formation rather than enhancement of
 the star formation.
  But at the same time, their simulation indicates that
 some cluster galaxies associated with the merging can survive the
 ram-pressure stripping and maintain their blue colors if their
 relative velocities to the cluster center are high, simply because
 they pass through the cluster too fast for the ram-pressure 
 to quench the star formation.
  Due to this effect, the relative velocities of blue galaxies are
 larger than those of red galaxies by about $\sim1000-1500$\,km s$^{-1}$
 in a merging cluster at 3.6 Gyr after the core-crossing. 
  The outcome of the simulation matches well with our results -- i.e.,
 no particular enhancement of star formation activities and the 
 peculiar distribution of strong/intermediate MIR-excess galaxies
 that are found preferentially at the blue velocity peak. 
  Therefore, we suggest that the discovery of the preferential 
 occupation of the blue velocity peak by strong/intermediate
 MIR-excess galaxies is a supporting evidence of the galaxy transformation
 taking place in a merging cluster following the picture presented in
 Fujita et al. (1999).

  These key findings allow us to draw a global picture of the galaxy 
 evolution process taking place in A2255. As witnessed from the
 strong/intermediate MIR-excess galaxies at the blue velocity peak
 and the lack of notable enhancement of star formation both globally
 in the cluster and in substructures, the cluster-cluster merging
 suppresses the star formation activity with the ram-pressure stripping
 rather than enhancing it. The quenching of star formation is 
 happening in two folds, one related to the cluster-cluster merging,
 and another related to the ram-pressure stripping of gas in 
 (infalling) galaxies at $0.3~R_{vir} < R_{cl} < R_{vir}$ where the
 supporting evidence comes from the spatial distribution of 
 intermediate MIR-excess galaxies and the lack of star-forming galaxies
 at $R_{cl}<0.5$\,Mpc.

\section{Conclusion}

  We investigated the MIR properties of galaxies in a merging cluster
 A2255 using the AKARI MIR observation over 3-24\,$\mu$m in addition
 to the \textit{Spitzer} MIPS 24, 70\,$\mu$m data, in order to understand the
 ongoing SFR activities in such a cluster, and the role of the
 environment on the subsequent evolution of the star-forming galaxies
 into red, quiescent galaxies.

  As a way to trace the evolutionary sequence of galaxies from the
 star-forming stage through the transition stage to the dead, quiescent
 stage, we examined the MIR colors ($N3-S11$ or $z-24~\mu$m) of the 
 cluster member galaxies. We found that MIR colors of the cluster
 galaxies show a large dispersion unlike the tight red of blue 
 sequences in the optical or NIR color-magnitude relation. 
  Virtually almost all ($>90$\%) of the member galaxies with MIR 
 detection have redder $N3-S11$ colors than the expectation from 
 stellar photospheric emission only. The MIR-excess can be categorized
 in three classes according to $N3-S11$ (or $z-24~\mu$m) color; 
 the first is a population of galaxies forming
 a blue sequence at $\langle N3-S11 \rangle \sim -1.7$
 (``weak'' MIR-excess) that can be considered as a coeval population with the  
 dust emission coming from the circumstellar dust around AGB stars.
  The second case is a population with a ``strong'' MIR-excess ($N3-S11 > 0.2$),
 most of which are star-forming galaxies with blue optical colors. The third case
 is an ``intermediate'' MIR-excess population ($-1.2<N3-S11<0.2$) which lie 
 between the weak and the strong MIR-excess galaxies. These are mostly passive
 galaxies with young mean stellar age.
  As such, the MIR color works as a classifier of galaxies at different
 stages of evolution.
  The intermediate MIR-excess galaxies have late-type morphology ($\sim80$\,\%),
 while the weak MIR-excess galaxies are predominantly early-type 
 ($\sim80$\,\%). The strong MIR-excess galaxies are mostly late-type galaxies.
 The morphologies of the different MIR-excess classes support the idea that
 the intermediate MIR-excess galaxies are the transition population bridging
 the strong and the weak MIR-excess galaxies. 

  Armed with the three different classes of MIR-excess galaxies which 
 trace different stages of the galaxy evolution, we addressed whether 
 the star formation is enhanced in the merging cluster or not, and 
 how the environment affects the galaxy transition. 
  Using star-forming galaxies as represented by strong MIR-excess galaxies, 
 we derived the total SFR of A2255.
  The IR luminosities of individual galaxies 
 range $6.0\times10^8~L_{\odot} < L_{IR} < 3.2\times10^{10}~L_{\odot}$,
 with the total SFR for entire cluster being $\sim130~M_{\odot}$ yr$^{-1}$. 
 The integrated SFR, and the IR luminosity function of A2255 is 
 consistent with those of other galaxy clusters at similar redshifts
 and with similar masses.
  This supports the idea that cluster scale dynamics 
 (cluster-cluster merging) does not enhance the star formation activity.
  Yet, it should be noted that most available 
 galaxy clusters studied to date are biased to dynamically unrelaxed 
 systems as A2255. 

  We identify substructures and two distinct velocity components
 (the blue velocity peak and the red velocity peak) in A2255. 
 To understand how large scale (cluster-cluster) and small scale
 (group/galaxy infall) merging activities affect the galaxy 
 evolution, the star formation and its quenching of MIR-excess galaxies
 were examined in these distinct components in A2255 as well as a function
 of the cluster-centric distance and the local density. 
  Star-forming galaxies are not only preferentially located in the 
 outer, lower density part of the cluster, but also found to be more
 abundant in the blue velocity peak. The latter fact has been predicted
 in simulations of merging cluster, where galaxies in a highest
 velocity component suffer less quenching from the ram-pressure
 stripping. No marked increase in the specific star formation is found
 over the cluster region. The relative fraction of intermediate 
 MIR-excess galaxies increases at the intermediate density region 
 or at regions with cluster-centric distance of 0.5-2.0\,Mpc,
 suggesting that quenching in star formation occurs over such a region
 where the ram-pressure stripping has been speculated to be a main
 star formation suppression mechanism. 
  These findings supports idea that the cluster-scale merging 
 suppresses the star formation and the ram-pressure stripping is 
 a main mechanism of the star formation quenching in A2255.

\acknowledgements

  We thank L. Piovan for providing his SED model. We would like to thank 
 R. Gobat, J. H. Lee for helpful discussions. This work is based on 
 observations with \textit{AKARI}, a JAXA project with the participation
 of ESA. The \textit{AKARI} images are obtained from a Mission Program 
 CLEVL (CLusters of galaxies EVoLution studies).
 This work was supported by the Korea Science and Engineering
 Foundation (KOSEF) grant No. 2009-0063616,
 funded by the Korea government (MEST).

\clearpage

\begin{deluxetable}{ccccc}
 \tabletypesize{\scriptsize}
 \tablecaption{\label{tab:obs_summary}
  Observational Parameters of A2255 Data }
 \tablewidth{0pt}
 \tablehead{
  \colhead{Filter} & \colhead{$\lambda_{eff}$} &
  \colhead{$t_{int}$}\tablenotemark{a} & \colhead{$5\sigma$ flux limit}\tablenotemark{b} &
  \colhead{FWHM}  \\
  \colhead{} & \colhead{($\mu$m)} &
  \colhead{(sec)} & \colhead{($\mu$Jy)} &
  \colhead{($\arcsec$)}
  }
\startdata
  N3  &  3.2 & 133.2 &  25 &  4.1 \\
  N4  &  4.1 & 133.2 &  30 &  4.0 \\
  S7  &  7.2 & 147.3 &  65 &  4.9 \\
  S11 & 10.4 & 147.3 &  80 &  5.3 \\
  L15 & 15.9 & 147.3 & 150 &  5.4 \\
  L24 & 23.0 & 147.3 & 400 &  6.3 \\ 
\enddata
\tablenotetext{a}{Total integration time per pixel}
\tablenotetext{b}{$5\sigma$ flux limits are measured within an aperture
 of $2\times$FWHM diameter in each band image.}
\end{deluxetable}

\begin{deluxetable}{cc cccccc cc cc}
 \tabletypesize{\scriptsize} 
 \tablecaption{\label{tab:phot} AKARI IRC 3--24\,$\mu$m, and \textit{Spitzer}
  MIPS 24/70\,$\mu$m photometry of Abell 2255 member galaxies  }
 \tablewidth{0pt}
 \tablehead{ 
  \colhead{ID}\tablenotemark{a} & \colhead{redshift}\tablenotemark{b} &
  \colhead{N3}\tablenotemark{c} & \colhead{N4} & \colhead{S7} & \colhead{S11} &
  \colhead{L15} & \colhead{L24} & \colhead{24\,$\mu$m} & \colhead{70\,$\mu$m} 
  }
\startdata  
A2255\_S171036.20+642003.0 & 0.077836 & 16.270 & 16.639 & 14.913 & 14.798 & $-1.000$ & $-1.000$ & 14.365 & 10.411 \\
A2255\_S171117.90+640800.0 & 0.082396 & 16.259 & 16.873 & 16.090 & 15.598 & 15.901 & 15.436 & 15.511 & 13.326 \\
A2255\_S171226.40+640456.0 & 0.079492 & 16.804 & 17.504 & 18.109 & 99.000 & 18.561 & 99.000 & 99.000 & 99.000 \\
A2255\_S171234.10+640550.0 & 0.076180 & 17.841 & 18.397 & 16.762 & 16.311 & 16.025 & 14.722 & 14.772 & 11.640 \\
A2255\_S171236.10+640508.0 & 0.082686 & 16.958 & 17.623 & 18.347 & 18.509 & 17.840 & 99.000 & 99.000 & 99.000 \\
A2255\_P171240.34+640443.1 & 0.083    & 17.248 & 17.932 & 17.743 & 17.135 & 17.777 & 99.000 & 17.962 & 12.882 \\
A2255\_P171247.18+635625.0 & 0.079    & 17.218 & 17.806 & $-1.000$ & $-1.000$ & 99.000 & 17.584 & 17.020 & 99.000 \\
A2255\_P171257.56+641028.2 & 0.087    & 18.090 & 18.629 & 17.412 & 16.715 & 17.281 & 16.967 & 16.166 & 13.146 \\
A2255\_P171335.98+640747.1 & 0.090    & 16.971 & 17.612 & 17.408 & 16.876 & 17.092 & 16.759 & 16.857 & 99.000 \\
A2255\_S171343.50+640502.0 & 0.090206 & 99.000 & 99.000 & 15.534 & 15.285 & 15.658 & 15.527 & 15.410 & 11.652 \\
A2255\_S171352.00+640710.0 & 0.082871 & 17.333 & 17.818 & 16.036 & 15.743 & 15.741 & 15.455 & 15.229 & 12.931 \\
A2255\_P171406.11+641019.9 & 0.089 & 17.570 & 18.205 & 17.383 & 17.079 & 17.554 & 17.328 & 17.314 & 99.000 \\
A2255\_P171452.86+640649.3 & 0.079 & 17.293 & 17.942 & 18.600 & 99.000 & 99.000 & 99.000 & 99.000 & 99.000 \\
A2255\_S171602.10+635729.0 & 0.079818 & $-1.000$ & $-1.000$ & $-1.000$ & $-1.000$ & 16.418 & 16.028 & 99.000 & 99.000 \\
A2255\_P171602.36+635755.0 & 0.079000 & $-1.000$ & $-1.000$ & $-1.000$ & $-1.000$ & 17.396 & 17.182 & 99.000 & 99.000 \\ 

\enddata
 \tablecomments{The complete version of this table is included in the electronic
  edition of the Journal. The printed edition contains only 15 objects as a sample. }
 \tablenotetext{a}{ID of galaxy represents whether the galaxy is a spectroscopic member 
  (`A2255\_S') or photometric member (`A2255\_P'), and the coordinates of the galaxy 
  in sexagesimal format in ra/dec.}
 \tablenotetext{b}{Redshifts of the spectroscopically selected members (`S') are 
  from SDSS DR2 \citep{Abazajian04}, while redshifts of the photometrically selected members 
  (`P') are from \citet{Yuan03}.}
 \tablenotetext{c}{The unit of N3, N4, S7, S11, L15, L24, MIPS 24\,$\mu$m and 70\,$\mu$m 
  values are AB magnitudes. The magnitudes represent total magnitudes of the galaxies.  
  The value $99.000$ indicates non-detection (below the detection limit- see text for details),
  and $-1.000$ indicates that the source is either out of the field of view in the
  observation, blended, etc. }
\end{deluxetable}

\begin{deluxetable}{l cccc cccc}
\tabletypesize{\scriptsize}
 \tablecaption{\label{tab:mirx_fraction} Number of A2255 Member Galaxies }
 \tablewidth{0pt}
 \setlength{\tabcolsep}{0.05in}
 \tablehead{
   \multicolumn{1}{c}{}  & \colhead{} & \multicolumn{3}{c}{S11}  & \colhead{} & \multicolumn{3}{c}{24\,$\mu$m}  \\
 \cline{3-5}  \cline{7-9} \\
   \colhead{} & 
   \colhead{} &
   \colhead{ $r<17.5$ } & \colhead{ $17.5<r<18.5$ } & \colhead{ $r>18.5$ } &
   \colhead{} & 
   \colhead{ $r<17.5$ } & \colhead{ $17.5<r<18.5$ } & \colhead{ $r>18.5$ }  }
\startdata
  Cluster member galaxies         & & 81 & 78 &     54 & & 158 & 146 & 122 \\ 
  MIR detections                  & & 63 & 26 &     13 & &  59 &  38 &  22 \\
  Galaxies with MIR-excess        & & 60 & 24 &     13 & &  59 &  38 &  22 \\ 
\tableline
  Weak MIR-excess                 & & 42 &  1 &$\ldots$& &  12 & $\ldots$ & $\ldots$  \\
  Intermediate MIR-excess         & &  9 &  8 &$\ldots$& &  20 &   3 & $\ldots$  \\
  Strong MIR-excess               & &  9 & 15 &     13 & &  27 &  35 &  22 
\enddata
\tablecomments{
  The columns 1-3 indicate the number of A2255 member galaxies in 
 AKARI/IRC S11 field of view, and the columns 4-6 indicate the number of 
 galaxies in \textit{Spitzer}/MIPS 24\,$\mu$m area coverage. 
  MIR detections are defined as $f_{S11} > 80~\mu$Jy and $f_{24~\mu m} > 250~\mu$Jy,
 $5\sigma$ flux limits as mentioned in Section 2. 
  The criteria for weak/intermediate/strong MIR-excess galaxies are 
 described in Section 3.2.  }
\end{deluxetable}

\clearpage

\end{document}